\documentclass[sigconf]{acmart}
\usepackage{accsupp}
\usepackage{graphicx}
\usepackage{array}
\usepackage{url}
\newcolumntype{C}[1]{>{\centering\arraybackslash}p{#1}}
\usepackage{comment}
\copyrightyear{2026}
\acmYear{2026}
\setcopyright{cc}
\setcctype{by}
\acmConference[CHI '26]{Proceedings of the 2026 CHI Conference on Human Factors in Computing Systems}{April 13--17, 2026}{Barcelona, Spain}
\acmBooktitle{Proceedings of the 2026 CHI Conference on Human Factors in Computing Systems (CHI '26), April 13--17, 2026, Barcelona, Spain}
\acmDOI{10.1145/3772318.3790405}
\acmISBN{979-8-4007-2278-3/2026/04}

\begin{document}

\title[Disability-First AI Dataset Annotation]{Disability-First AI Dataset Annotation: Co-designing Stuttered Speech Annotation Guidelines with People Who Stutter}
\author{Xinru Tang}
\orcid{0000-0001-6426-1363}
\affiliation{\institution{University of California, Irvine}\city{Irvine}\state{California}\country{USA}}
\email{xinrut1@uci.edu}
\affiliation{\institution{AImpower.org}\city{Mountain View}\state{California}\country{USA}}
\email{xinru@aimpower.org}

\author{Jingjin Li}
\orcid{0000-0003-0193-7228}
\affiliation{\institution{AImpower.org}\city{Mountain View}\state{California}\country{USA}}
\email{jingjin@aimpower.org}

\author{Shaomei Wu}
\authornote{Corresponding author}
\orcid{0000-0003-1104-4116}
\affiliation{\institution{AImpower.org}\city{Mountain View}\state{California}\country{USA}}
\email{shaomei@aimpower.org}

\begin{abstract}
Despite efforts to increase the representation of disabled people in AI datasets, accessibility datasets are often annotated by crowdworkers without disability-specific expertise, leading to inconsistent or inaccurate labels. This paper examines these annotation challenges through a case study of annotating speech data from people who stutter (PWS). Given the variability of stuttering and differing views on how it manifests, annotating and transcribing stuttered speech remains difficult, even for trained professionals. Through interviews and co-design workshops with PWS and domain experts, we identify challenges in stuttered speech annotation and develop practices that integrate the lived experiences of PWS into the annotation process. Our findings highlight the value of embodied knowledge in improving dataset quality, while revealing tensions between the complexity of disability experiences and the rigidity of static labels. We conclude with implications for disability-first and multiplicity-aware approaches to data interpretation across the AI pipeline.
\end{abstract}

\ccsdesc[500]{Human-centered computing~Empirical studies in accessibility}
\ccsdesc[500]{Human-centered computing~Accessibility design and evaluation methods}
\ccsdesc[300]{Computing methodologies~Artificial intelligence}

\keywords{AI dataset annotation, accessibility, disability, stuttering, embodied knowledge}

\maketitle
\section{Introduction}
Recent advances in artificial intelligence (AI), such as large language models (LLMs) and other generative AI models, have relied on the use of vast amounts of human data~\cite{bender-bigLLM}. Meanwhile, a growing body of work has documented AI models' degraded, and at times discriminatory, performance for users with disabilities\footnote{We use both people-first (people with disabilities) and identity-first (disabled people) language to recognize diverse naming preferences within disabled communities.}~\cite{massiceti2024explaining, charan2025whisper, whittaker-2019, hutchinson-2020, gadiraju2023wouldn,zhao-facerec-2018, koenecke-aphasia-2024}, and identified the lack of data representing disabled people as one of the root causes~\cite{whittaker-2019, bragg2021fate, li2024towards, hutchinson-2020, park2021designing, sharma2023disability}. In response, researchers have sourced data from disabled people to create \textit{accessibility datasets}~\cite{kacorrie-2020-incluset, desai2023asl, li2024want, gong202470, gurari_vizwiz_2018, sharma2023disability, massiceti2021orbit}. Within this movement, the notion of ``disability-first datasets'' was proposed to emphasize the importance of centering disabled communities' experiences and interests in both the creation and application of accessibility datasets ~\cite{theodorou2021disability, sharma2023disability,li2024want}. 

Despite ongoing data collection efforts, label noise in accessibility datasets remains a persistent yet underexplored problem. One notable example is VizWiz, an accessibility dataset of photos taken by blind and low-vision (BLV) individuals for visual question answering~\cite{gurari_vizwiz_2018}. Prior work reported over 40\% disagreement between crowdworkers and in-house experts on whether a photo contained privacy or quality issues~\cite{gurari_vizwiz_2018}, and only 9\% unanimous agreement among annotators on the answer to associated visual questions~\cite{gurari-2017-crowdverge}. Similar label inconsistencies have been observed in datasets of sign languages ~\cite{bragg2021fate} and stuttered speech~\cite{lea2021sep, charan2025whisper}. 

Two main factors may have contributed to these data labeling issues: (1) the distinctive characteristics of data sourced from disabled people (e.g. blurriness in photos taken by BLV people ~\cite{gurari_vizwiz_2018}), and (2) annotators' limited understanding of disabled people's needs and experiences~\cite{gurari-2017-crowdverge, charan2025whisper, li2024want, simons2020hope}. While AI data annotation tasks have become increasingly interpretive~\cite{wan2025noise, cambo2022model, scheuerman2025data, scheuerman2020we}, most accessibility datasets are still annotated by data workers without relevant experience or expertise in disabilities to properly interpret the data. For instance, VizWiz was annotated by sighted crowdworkers~\cite{gurari_vizwiz_2018}, who reportedly were uncertain about the kinds of visual information most useful to BLV people ~\cite{simons2020hope}. Similarly, Sep-28k, the largest English stuttered speech dataset sourced from stuttering related podcasts and labeled by annotators ``\textit{who received training via written descriptions, examples, and audio clips on how to best identify each dysfluency but were not clinicians}''~\cite{lea2021sep}, was observed by researchers who stutter to mislabel a significant amount of natural disfluencies (e.g. fillers, pauses) as stuttering disfluencies~\cite{charan2025whisper}\footnote{Authors of Sep-28k did not specify whether the annotators stutter or not~\cite{lea2021sep}. We can only assume the annotators are not PWS as such information would have been provided if they are.}.

As an initial step toward addressing these issues, our study explores a disability-first approach to accessibility dataset annotation through a case of transcribing and annotating stuttered speech. Speech disfluencies, whether natural or stuttering-induced, are typically omitted from transcriptions in existing speech AI  datasets~\cite{radford2022robustspeechrecognitionlargescale}, resulting in the \textbf{default erasure} of stuttering by automatic speech-to-text systems~\cite{li2024want, li-collective-2025}. Even when stuttering is annotated, inter-rater agreement remains alarmingly low. For instance, Sep-28k reported an inter-rater agreement of 0.11 for identifying speech prolongations, 0.25 for blocks, and 0.39 for determining whether a clip contains disfluencies~\cite{lea2021sep}. While annotating stuttered speech is inherently challenging and subjective due to its high variability and the ambiguous boundary between stuttered and fluent speech~\cite{tichenor2019stuttering, valente2025clinical}, people who stutter (PWS) are rarely involved to address such complexities~\cite{batra2025boli, lea2021sep, ratner2018fluency, valente2025clinical}. Building on a recent shift in clinical research that prioritizes the subjective experience of stuttering over listener perspectives~\cite{constantino2020speaker}, our study aims to incorporate the embodied knowledge and lived experiences of PWS to develop more reliable annotations of stuttered speech for AI.

More specifically, our study sets out to co-design stuttered speech annotation guidelines with PWS. Following the disability-first idea ~\cite{theodorou2021disability}, we developed, evaluated, and evolved our guidelines by centering the stuttering experience and identity, rather than ``co-opting'' existing speech annotation for so-called ``edge cases'' like stuttering~\cite{motepalli23_interspeech}. To do so, we crafted our co-design process into three phases: 1) \textit{Formative studies} in which we reviewed and discussed the annotation practices of existing stuttered speech datasets with AI professionals who stutter (PWS AI professionals); 2) \textit{Co-design sessions} in which we designed, tested, and iterated a new set of stuttered speech annotation guidelines with PWS AI professionals and speech-language pathologists (SLP) specializing in stuttering; 3) \textit{Evaluation sessions} in which we reviewed the resulting annotations of their own speech with participants who stutter. 

Our experiences show that involving PWS in data annotation is crucial for representing stuttered speech in ways that align with the lived experience of stuttering, as they bring lived, deeply embodied experiences of stuttering that fluent speakers typically do not fully understand. For instance, participants who stutter shared how they attended and leveraged non-verbal acoustic cues (e.g. change in breathing patterns) to identify and distinguish different types of stuttering. Our findings also suggest the inherent subjectivity of stuttering annotation: even PWS and SLP clinicians with extensive exposure to stuttering often perceive stuttered speech differently, and acoustic signals alone could be fundamentally limited in capturing the multitudes and complexity of stuttering.

Our study makes three main contributions to the HCI literature on AI for accessibility.
\begin{enumerate}
    \item \textit{Methodological.} Extending existing accessibility data research that has primarily focused on data collection~\cite{desai2023asl, gurari_vizwiz_2018, kamikubo2024accessshare, kamikubo2021sharing, theodorou2021disability, sharma2023disability}, we explore a disability-first approach that enables PWS to shape the interpretation of data from the outset and throughout the annotation process.
     
    \item \textit{Artifact.} The resulting guidelines from our co-design study represent the first effort to develop PWS-centered stuttered speech annotation guidelines for speech AI development (see Appendix \ref{appendix::guide}). In contrast to conventional annotation guidelines that focus primarily on audible speech ~\cite{ratner2018fluency, lea2021sep}, our guidelines draw on the lived experiences of PWS to highlight the importance of non-verbal signals (e.g. breathing, pauses) in communication, and caution against the common practice of trimming non-speech audio during automatic speech recognition (ASR) model training~\cite{lamel-2002-trim}.  
    
    \item \textit{Empirical.} Our study identifies underexplored challenges in annotating stuttered speech (see Appendix \ref{appendix::challenges}), with broader implications for annotating accessibility datasets. As illustrated by stuttering, disability experience is inherently dynamic and situated, defying the static, categorical, and often binary framings typically imposed in AI data annotation. We therefore urge the AI research community to recognize the embodied knowledge of disabled people and to embrace greater multiplicity and interpretive flexibility when engaging with disability data throughout the AI pipeline.
\end{enumerate}
 
\paragraph{Positionality}
This study is shaped by our positionality as researchers working at the intersection of HCI, accessibility, machine learning, and speech technologies. Our research team includes both PWS (Wu) and non-PWS members (Tang and Li), with the PWS  member serving as the senior author and guiding the research direction. Our relationships with the stuttering community are not uniform and have shaped how we approached this work. Tang entered the topic as a relative newcomer to stuttering; however, she brought prior experience in participatory and disability-centered research~\cite{tang2023community, tang2025everyday, tang2026reimagining, garg2025s, tran2026toward}, which informed her approach to listening and collaborative decision-making. Li and Wu have long-standing involvement with the stuttering community, including efforts to develop stuttering-friendly  technologies~\cite{wu2023world, wu2024autoethnography, li2024towards, li2024want, li-2024-codesign, li-collective-2025, li2025govern} and to support stuttering advocacy. Li and Wu also hold deep personal and professional ties within the community, which have influenced not only recruitment and design decisions but also our interpretation of the values, tensions, and priorities surfaced throughout this research. Collectively, our understandings of stuttering and disability are informed by our personal experiences and academic training within U.S.-based institutions, such as disability activism, critical disability studies, the neurodiversity movement, and the affirmative approach in stuttering research and therapy (e.g.,~\cite{constantino2018can, constantino2020speaker, constantino2023fostering}). We acknowledge the diversity of experiences and orientations within the stuttering community and have tried our best to respect such diversity throughout our study.

\section{Related Work}
\subsection{Stuttering as a Contested Disability}
Stuttering is a neurodevelopmental condition characterized by involuntary disfluencies in speech, affecting over 1\% of the global population ~\cite{stuttering}. The field of SLP has historically framed stuttering as a problem to be fixed, relying on external observations and control ~\cite{constantino2020speaker, tichenor2019stuttering}. Yet, the measurement of stuttering has been a contested issue in SLP due to its variability and the wide range of associated behaviors ~\cite{yaruss1997clinical}. Patterns and frequency of stuttering differ not only across individuals but also within the same speaker over time and in response to changing contexts, such as task, setting, or conversational partner ~\cite{yaruss1997clinical, tichenor2021variability}. As a result, having an operationalized definition of stuttering has proven to be a difficult task in SLP, and currently there is no universally agreed-upon definition ~\cite{stuttering-revisited}. Annotating stuttered speech is found to be highly subjective task even for SLP professionals ~\cite{valente2025clinical}.

Pushing back on the medical model of stuttering, an increasing number of voices from the stuttering community have called for the inclusion of PWS in shaping knowledge about stuttering ~\cite{constantino2020speaker,tichenor2019stuttering} as well as the prioritization of PWS's perspectives and lived experience in stuttering research and therapy~\cite{constantino2022stuttering, constantino2023fostering, constantino2020speaker, tichenor2019stuttering}. Constantino contends that spontaneity, rather than fluency, should be the guiding objective of speech therapy ~\cite{constantino2020speaker}. Other research further emphasizes the affective, behavioral, and cognitive dimensions that shape the lived experience and intimate understanding of stuttering ~\cite{tichenor2019stuttering}. Our work contributes to these ongoing efforts by centering the lived experiences of PWS in data annotation as a central way of shaping speech AI technologies.

\subsection{Disability-First AI Dataset Creation}
Theodorou et al. proposed \emph{disability-first} as an approach ``\textit{serving a disability community first},'' which ``\textit{stands in opposition to mainstream ML datasets and approaches which are later augmented or co-opted to address issues of importance to disabled communities}'' ~\cite{theodorou2021disability}. Reflecting similar ideas and in response to general-purpose datasets that often fail to adequately represent or prioritize disabled people's perspectives and needs~\cite{cao2022s, charan2025whisper, li2024towards}, researchers have created disability-first datasets such as VizWiz ~\cite{gurari_vizwiz_2018}, Orbit~\cite{massiceti2021orbit}, ASL Citizen ~\cite{desai2023asl}, and AS-70 ~\cite{gong202470, li2024want} -- all of which directly sourced from disabled communities. Other work has explored collection mechanisms aimed at supporting disabled people as data contributors, including accessible designs of privacy access control for blind users ~\cite{kamikubo2024accessshare}, community-led data collection method for stuttered speech~\cite{li2024want}, and data anonymization tools designed to protect deaf\footnote{Deaf communities in the U.S. often capitalize `D' in `Deaf' to emphasize a shared cultural identity. We do not differentiate between Deaf and deaf in our writing, as this distinction is increasingly contested within deaf studies ~\cite{kusters2017innovations}. We use deaf to acknowledge the fluidity of identity and to recognize that access to deaf cultural resources itself can be a form of privilege. We use deaf and hard-of-hearing (DHH) to encompass broader populations who have hearing disabilities.} users in sharing sign language videos ~\cite{bragg2022exploring}. However, these efforts face ongoing challenges, including open questions around fair compensation, consent, and ethical engagement with disabled communities ~\cite{park2021designing}.

Despite recent disability-first data collection efforts, a critical gap remains in data annotations. Disabled people are typically treated solely as the source of data during data collection, excluded from other crucial stages of the AI development pipeline such as data annotation and model design~\cite{li2025govern}. Annotation of accessibility datasets often relies on crowd workers who lack familiarity with the target communities. VizWiz, one of the largest and most widely used visual accessibility datasets sourced from BLV people, relies on sighted crowd workers to provide labels and determine label accuracy through inter-annotator agreement ~\cite{gurari_vizwiz_2018}. However, sighted crowd workers often apply inconsistent interpretive standards shaped by their own assumptions and biases ~\cite{simons2020hope}. These inconsistencies can significantly impact benchmarking and the usability of downstream applications for BLV users ~\cite{kapur2024reference, garg2025s}. Similar limitations are common in accessibility datasets, such as sign languages ~\cite{desai2024systemic, bragg2021fate} and stuttered speech ~\cite{lea2021sep}, which rarely involve affected communities in the annotation process.

These issues in accessibility dataset annotation align with broader calls in critical AI research to shift attention to the underlying systems of power in AI development ~\cite{miceli2022studying, barabas2020studying}. Prior research has shown that so-called `ground truth' labels are shaped by specific contexts, such as annotators' backgrounds and expertise, as well as broader organizational control ~\cite{muller2021designing, abdelkadir2025role}. For example, studies of industry-scale data collection revealed that the pursuit of label precision is often driven by market-oriented values such as objectivity, standardization, and corporate interests ~\cite{zhang2025making, miceli2020between, kazimzade2020biased, zhang2024conceptualizing, kapania2023hunt}. Other research has uncovered the ambiguity and uncertainty inherent in socially constructed labels, such as gender ~\cite{scheuerman2020we}, accent ~\cite{prinos2024speaking}, and content toxicity ~\cite{cambo2022model, abdelkadir2025role}. Consequently, a growing body of scholarship emphasizes the need to document the production processes and contextual factors involved in dataset creation ~\cite{miceli2021documenting, miceli2022documenting}. Research has also called for more reflective data practices, advocating for the treatment of bias or error as a site of negotiation rather than simply as model failures ~\cite{cambo2022model, lin2023bias}. Our work extends this line of research to the domain of accessibility and disability through a case study of stuttered speech annotation.

\BeginAccSupp{method=pdfstringdef,unicode,Alt={A flowchart divided into three columns, representing the three phases of a study. The columns are labeled "Phase 1," "Phase 2," and "Phase 3."
- Phase 1, "Formative Studies," is on the left. It shows a panel with texts and two icons at the top: one of datasets review and another of interviews with 2 PWS AI professionals. Below these is a section labeled "Takeaways" listing two points.
- Phase 2, "Co-design Sessions," is in the middle. It shows a panel with an arrow points from a pilot study with "3 PWS AI Professionals" to an iterative study with "2 SLP Professionals." Below this is a "Takeaways" section with two points.
- Phase 3, "Evaluation Sessions," is on the right. It shows a panel texts with an icon of interviews with 4 speech data contributors. Below this is a "Takeaways" section with two points.}}
\begin{table*}[!t]
\resizebox{\linewidth}{!}{
\begin{tabular}{
l   
l   
l   
l  
l 
l  
p{3.0cm}   
l   
}
\hline
\textbf{Dataset}    & \textbf{Field} & \textbf{Speakers}  & \textbf{Tasks} & \textbf{Transcripts} & \textbf{Annotations} & \textbf{Annotators} & \textbf{Languages} \\ \hline
FluencyBank* ~\cite{ratner2018fluency}        & SLP            & \begin{tabular}[c]{@{}l@{}}22 CWS \& \\ 38 AWS\end{tabular}             & \begin{tabular}[c]{@{}l@{}}conversation, \\ reading\end{tabular}                                                 & Yes                    & N/A                                                                                     & N/A                                                                      & English                                                                                          \\ \hline
UCLASS* ~\cite{howell2009university}             & SLP            & 25 CWS                                                                        & conversation                                                                                                     & \begin{tabular}[c]{@{}l@{}}Yes\end{tabular}                 &  N/A  & N/A                                                                      & English                                                                                          \\ \hline
KSoF ~\cite{bayerl2022ksof}                & SLP            & 37 PWS                                                                        & \begin{tabular}[c]{@{}l@{}}reading, \\ spontaneous speech \\ (e.g., dialogues, \\ scene description)\end{tabular} & No                     & clip-level                                                                              & non-PWS students                                                                 & German                                                                                           \\ \hline
LibriStutter ~\cite{kourkounakis2021fluentnet}        & ML             & 
50 non-PWS**
& audiobook                                                                                                        & Yes                    & clip-level                                                                              & not reported                                                                     & English                                                                                          \\ \hline
AS-70 ~\cite{gong202470}               & ML             & 72 AWS                                                                        & \begin{tabular}[c]{@{}l@{}}conversation, \\ voice commands\end{tabular}                                          & Yes                    & word-level                                                                              & \begin{tabular}[c]{@{}l@{}}non-PWS professional\\ annotators; reviewed by\\ non-PWS supervisors \\ and a PWS researcher \end{tabular} & Mandarin                                                                                         \\ \hline
Boli ~\cite{batra2025boli}                & ML             & 28 PWS                                                                        & \begin{tabular}[c]{@{}l@{}}reading, \\ scene description \\ (spontaneous)\end{tabular}                           & Yes                    & syllable-level                                                                          & not reported                                                                     & \begin{tabular}[c]{@{}l@{}}English, Hindi, \\ Telugu, Bengali, \\ Marathi, Assamese\end{tabular} \\ \hline
Sep-28k ~\cite{lea2021sep}    & ML    & PWS podcasts                                                         & podcast                                                                                                 & No            & clip-level                                                                     & trained non-PWS                                                            & English                                                                                 \\ \hline
Sep-28k-SW ~\cite{charan2025whisper} & ML    & PWS podcasts                                                         & podcast                                                                                                 & Yes           & syllable-level                                                                & PWS researchers                                                         & English                                                                                 \\ \hline
\multicolumn{8}{p{\linewidth}}{* Limited to the transcribed portion of the dataset.\newline 
** Synthetic stutters were injected into fluent speech. \newline
\textit{Abbreviations}: AWS -- adults who stutter; CWS -- children who stutter; PWS -- people who stutter.}   \\
\end{tabular}
}
\caption{Examples of major stuttered speech datasets developed in SLP and ML research.}
\label{ref::table-datasets}
\end{table*}

\begin{figure*}[!t]
\centering
\includegraphics[width=17cm]{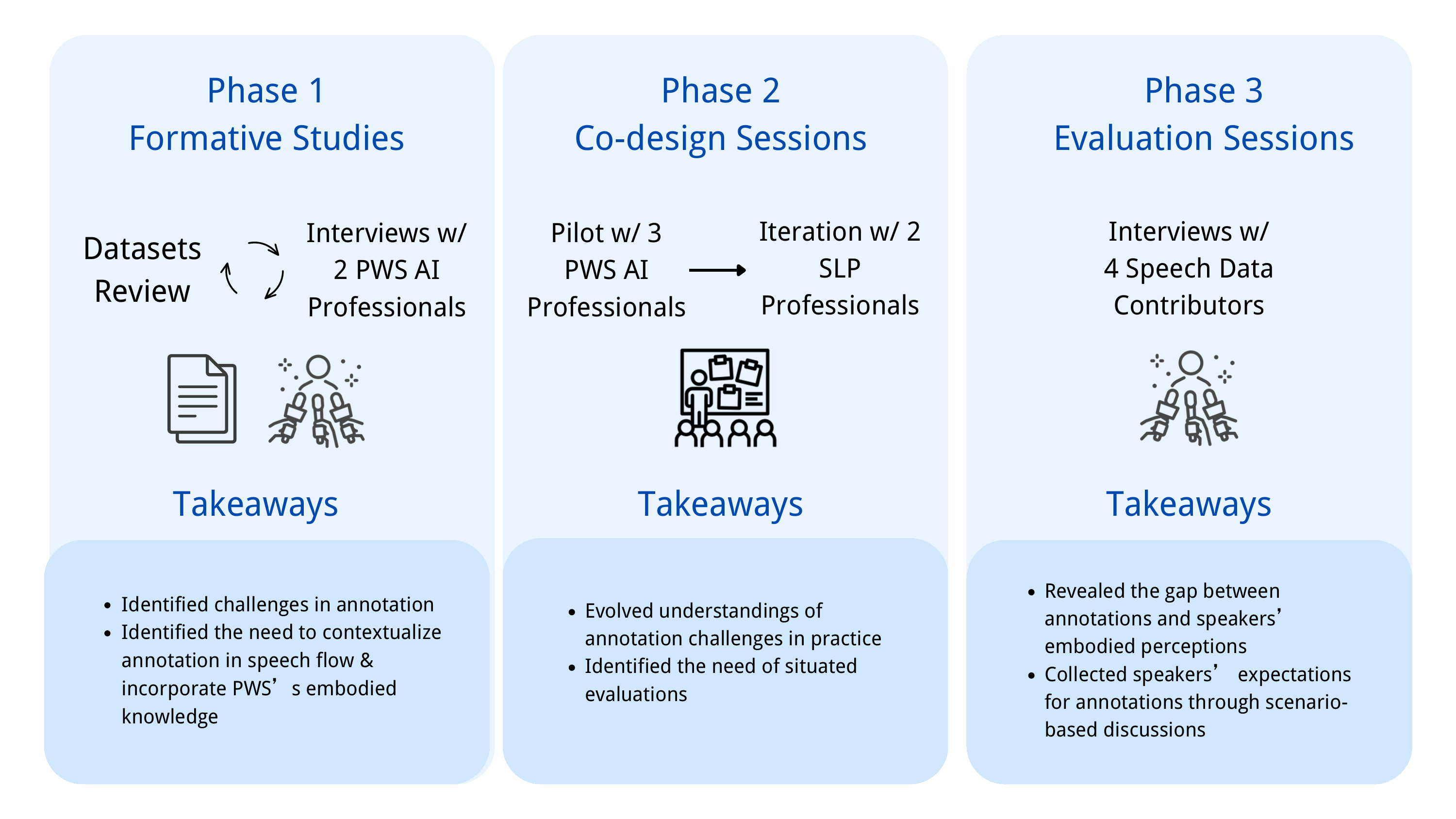}
\caption{Study flow with main takeaways at each stage.}
\label{fig::flowchart}
\end{figure*}
\EndAccSupp{}

\subsection{Stuttered Speech Datasets}
\label{sec::datasets-rw}
Existing stuttered speech datasets have been developed primarily for two purposes: corporate initiatives and academic research. Much recent development has been driven by corporate interest in ASR and a growing recognition of the need for more diverse speech data~\cite{macdonald2021disordered, zhang2022stutter, lea2023user, lea2021sep}. However, access to these datasets is often restricted, as they typically involve sensitive or proprietary data, except when sourced from public domains such as podcasts ~\cite{lea2021sep}. As a result, current research has mainly relied on publicly available datasets collected for academic use.

Datasets collected by academia primarily come from two fields: SLP and machine learning (see Table \ref{ref::table-datasets} for an overview). As training materials for speech pathologists, SLP datasets are typically limited in size. In contrast, datasets from the ML community tend to be more scalable, but still limited by annotation granularity and representativeness. For instance, as the largest publicly available English stuttering dataset, Sep-28k does not have transcriptions and provides only clip-level annotations rather than word-/syllable-level labels ~\cite{lea2021sep}. LibriStutter,  another influential English stuttering dataset, did not collect speech from PWS but injected synthetic stuttering sounds into read speech from fluent speakers ~\cite{kourkounakis2021fluentnet}. 

An underlying concern across these efforts is the limited power that PWS have in the creation of stuttered speech datasets. While PWS have frequently contributed as data sources, their influence on other critical aspects such as annotation frameworks, labeling guidelines, and the interpretation of stuttering behaviors remains minimal. For example, Sep-28k consists of over 28k three-second audio clips curated from public podcasts featuring PWS; however, the annotators' backgrounds remain unclear. Among the five stuttering labels it adopts, inter-annotator agreement scores were moderate for word repetitions (0.62) and interjections (0.57), but dropped significantly for sound repetitions (0.40), no disfluencies (0.39), blocks (0.25), and prolongations (0.11)~\cite{lea2021sep}. Furthermore, a recent study conducted by a team of PWS researchers (Sridhar and Wu) observed a significant number of ``mistakes'' in the original event labels and adjusted the labeling for over 25\% of the 2,621 clips in their sample for their benchmarking studies ~\cite{charan2025whisper}. Their work calls into question the original labels of Sep-28k and yields a re-annotated subset, which we hereafter refer to as Sep-28k-SW. An exception to these efforts is AS-70, where the quality control team included one PWS member who reviewed the annotations. However, AS-70 focused on Mandarin and the original annotations were still generated by non-PWS crowdworkers. Our work extends this line of research to English and centers PWS perspectives from the outset.

\section{Study Overview}
This study is part of a larger, ongoing initiative to develop representative speech datasets with and for PWS to make speech AI models more inclusive of stuttering. Prior to the present study, we have recorded conversational and reading speech from 51 English-speaking PWS, following a disability-first process similar to what was described in~\cite{li2024want}. The present study was directly motivated by labeling issues identified in existing stuttered speech datasets~\cite{lea2023user, charan2025whisper}. As an alternative to conventional annotation approaches, our work explores the development of PWS-led guidelines to more accurately and authentically represent stuttering in transcription and annotation practices.
Our study consists of three stages, as shown in Figure \ref{fig::flowchart}.
\begin{enumerate}
    \item \textit{Phase 1: Formative Studies.} We first reviewed and analyzed existing annotated datasets to identify gaps and limitations. Specifically, we examined different annotated versions of the Sep-28k dataset~\cite{lea2021sep, charan2025whisper} and interviewed two PWS AI professionals experienced in stuttered speech annotation to gather feedback on current practices and opportunities for improvement. We focused on Sep-28k due to its widespread use\footnote{Over 170 citations since its release in 2021 according to Google scholar (\url{https://scholar.google.com/scholar?cites=10057318411144677061}, accessed on 1/27/2026).}, its public availability of annotations produced by PWS~\cite{charan2025whisper} and non-PWS~\cite{lea2021sep}, and the reported discrepancy between these two versions~\cite{charan2025whisper}.
    At the conclusion of this phase, we drafted preliminary annotation guidelines to serve as the foundation for subsequent co-design sessions.
    \item \textit{Phase 2: Co-design.} We refined, piloted, and iterated our guidelines with PWS AI professionals and SLPs  specializing in stuttering. Together, we applied the preliminary guidelines to annotate stuttered speech samples, reflected on our experience and challenges in the application, and incorporated PWS participants' embodied knowledge of stuttering into the guidelines to mitigate challenges and better represent stuttering experiences. 
    \item \textit{Phase 3: Evaluation.} We evaluated the refined guidelines by reviewing the annotation of participants' own speech with four PWS data contributors.
\end{enumerate}

All study sessions were conducted over Zoom with participants' consent to record for analysis. As all but one participant were PWS (see Table~\ref{table::participants}), we adopted proactive strategies to create a supportive communication environment, including confirmation of turn taking, reserving buffer time to reduce time pressure, and encouraging the use of non-verbal channels such as emojis, body gestures, chat, and virtual whiteboard. Recognizing that listening to one's own stuttered speech can be uncomfortable~\cite{wu2023world}, we ensured participants were fully informed and explicitly consented before playing the recordings of their speech during the evaluation phase.

This study is exploratory and interpretative in nature. Our goal is to examine how the lived experiences of stuttering could inform and improve the annotation of stuttered speech. Therefore, aside from the descriptive statistical analysis conducted during our review of stuttered speech datasets in the formative stage, we employed reflexive thematic analysis as our primary method~\cite{braun2021can}. The data used in analysis involved notes from dataset reviews, transcripts from interviews, and video recordings of co-design workshops, all collected with participants' informed consent. Our analysis includes iterative and ongoing theme development along with data collection based on patterns of shared meaning among the data ~\cite{braun2021can}. We followed the same approach for all our qualitative data analysis. The lead author conducted open inductive coding of the cumulative data and regularly discussed the themes within the research team. The discussions centered on the distinct perspectives participants contributed to the transcription and annotation of stuttered speech for AI, as well as the challenges and trade-offs involved. We iterated on the coding process until reaching consensus over all the themes. This sole-coder approach aligns with the interpretative nature of qualitative research, and senior researchers were involved at all stages to enhance the reliability of the findings~\cite{campbell2013coding}.

\section{Formative Studies}
We began with formative studies consisting of two parts: (1) an iterative analysis of Sep-28k ~\cite{lea2021sep} and Sep-28k-SW ~\cite{charan2025whisper}; and (2) interviews with two PWS AI professionals who had annotated stuttered speech for their work.

\subsection{Methods}
\subsubsection{Datasets Review}
\begin{table}[t]
\begin{tabular}{p{0.35\columnwidth}|p{0.56\columnwidth}}
\hline
\textbf{Stuttering Events} & \textbf{Definitions}                                           \\ \hline
prolongations         & Elongated syllable: “M{[}mmm{]}ommy”                   \\ \hline
block                & Gasps for air or stuttered pauses                     \\ \hline
sound repetition       & Repeated syllables: \newline “I {[}pr-pr-pr-{]}prepared dinner” \\ \hline
word/phrase repetition    & Repeated words: \newline “I made {[}made{]} dinner”                            \\ \hline
interjection          & Filler words, e.g., “um”, “you know”          \\ \hline
\end{tabular}
\caption{Definitions of stuttering events in Sep-28k ~\cite{lea2021sep}.}
\label{Sep-28k-labels}
\end{table}

To explore annotation inconsistencies, we compared Sep-28k (the original version produced by crowdworkers ~\cite{lea2021sep}), and Sep-28k-SW (a re-annotated subset by Sridhar and Wu, both of whom self-identified as PWS ~\cite{charan2025whisper}). The audio recordings in Sep-28k are broken into 3-second clips and annotated by three annotators with five binary labels: word repetition, sound repetition, block, prolongation, and interjection (see Table \ref{Sep-28k-labels} for definitions) ~\cite{lea2021sep}. Sep-28k-SW was developed based on Sep-28k to benchmark ASR performance for stuttered speech~\cite{charan2025whisper}. Sridhar re-annotated a subset of Sep-28k, prioritizing clips with unanimous annotator agreements in the original Sep-28k. They reported making adjustments to over 25\% of the labels -- mostly due to the confusion between natural and stuttering disfluencies, and called for ``\textit{close attention to the content, flow, and voice quality}'' during stuttering annotation~\cite{charan2025whisper}. The lead author reviewed both datasets between March and June 2025. She identified and listened to audio clips with different labels\footnote{As each clip has only one single label in Sep-28k-SW but three labels in Sep-28k by three different annotators, we determined the final labels for Sep-28k using majority voting.} between Sep-28k and Sep-28k-SW, and documented her thoughts on the difference alongside each clip for further analysis. 

\begin{table*}[!ht]
\centering
\resizebox{\linewidth}{!}{
\begin{tabular}{
    p{0.09\linewidth}
    p{0.10\linewidth}
    p{0.15\linewidth}
    p{0.31\linewidth}
    p{0.33\linewidth}
}
\toprule
\textbf{Participant} & 
\textbf{Phase} &
\textbf{Background} &
\textbf{Community Activity} &
\textbf{Relevant Experiences} \\
\midrule

Charan & 
Formative; Co-design &
PWS, student, \newline AI researcher &
  
Host of stuttering self-help groups; volunteer at stuttering support organizations.   &
Used and annotated Sep-28k for research~\cite{charan2025whisper}. Co-author of Sep-28k-SW dataset. \\
\hline

Rong &
Formative; Co-design &
PWS, \newline tech worker in AI fields & 
Co-founder and executive director or StammerTalk community.  &
Used Sep-28k for work. Co-creator of Mandarin Stuttered Speech dataset~\cite{li2024want} (designed annotation guidelines; revised annotations by non-PWS). Published authors on speech AI for stuttered speech~\cite{gong202470, li-collective-2025}.  \\ 
\hline

Jia & 
Co-design &
PWS, \newline SLP professional &
Organizer and host of multiple stuttering self-help groups; co-founder, director, and/or advisor for multiple stuttering community organizations  &
Annotated stuttered speech in educational and clinical settings \\
\hline

Kerrigan &
Co-design &
non-PWS, \newline SLP professional &
Director and programming associate for multiple stuttering community organizations  &
Annotated stuttered speech in educational and clinical settings \\
\hline

Adedotun &
Evaluation &
PWS, \newline medical student & Volunteer at stuttering support organizations
&
Contributed speech to dataset \\ \hline

Amina &
Evaluation &
PWS, \newline PhD student &
 &
Contributed speech to dataset; prior experience transcribing own speech \\ \hline

Tatianna &
Evaluation &
PWS, \newline small business owner &
 &
Contributed speech to dataset \\ \hline

Benji &
Evaluation &
PWS, \newline tech worker in AI fields & Volunteer at stuttering support organizations
 &
Contributed speech to dataset; used multiple stuttered speech datasets for personal projects; deep technical expertise in speech AI. \\ \hline                  
\end{tabular}}
\caption{Participants' backgrounds. All of our participants in the Formative and Co-design phases have extensive experience organizing and leading stuttering community activities and advocacy efforts. All participants in the Evaluation Phase have previously contributed speech data and were asked to evaluate the annotation of their own speech. All participants have given explicit permission to use their real names.}
\label{table::participants}
\end{table*}

\subsubsection{Interviews}
Alongside our dataset review, we interviewed two PWS AI professionals (Charan and Rong in Table \ref{table::participants}) in March 2025 to incorporate their perspectives as PWS speakers and expertise in stuttered speech annotation. Besides being PWS, both have worked with English stuttered speech datasets and had extensive experience annotating stuttered speech. We recruited them from our personal network and received their informed consent to conduct audio/video recorded interviews for this study. 

The interviews were semi-structured and guided by an interview protocol including questions about their understanding of stuttering, experiences with annotation work, etc. (see Appendix \ref{appendix::interview-guide}). Both interviews were held over Zoom and each lasted about one hour. To facilitate reflection, we selected one stuttered speech audio sample in each participant’s native language from two publicly available stuttered speech datasets ~\cite{lea2021sep, gong202470} and invited them to explain how they considered the transcriptions and interpreted the stuttering event labels. We transcribed the interview audio recordings for analysis. The analysis resulted in four themes on challenges in annotations as presented in \ref{section::formative-findings}.

\subsection{Findings}
\label{section::formative-findings}
\begin{table}[]
\resizebox{\linewidth}{!}{\begin{tabular}{lrrcr}
\hline
\textit{"Does the clip contain ...?"}  & \textbf{FN} & \textbf{FP} & \textbf{total} & \textbf{\% in Sep-28k-SW}  \\ \hline
prolongation     & 35       & 288      & 323 & 12.32\% \\ \hline
block          & 67      & 251     & 318  & 12.13\%  \\ \hline
sound repetition  & 110  & 119     & 229  &  8.74\% \\ \hline
word repetition   & 55   & 88      & 143  &  5.46\%  \\ \hline
interjection     & 85    & 193     & 278   & 10.61\% \\ \hline
\end{tabular}}
\caption{Distribution of annotation disagreements between Sep-28k ~\cite{lea2021sep} and Sep-28k-SW ~\cite{charan2025whisper}. Treating the binary labels in Sep-28k-SW as \textit{actual} labels and the ones in Sep-28k as \textit{predictions}, FN (false negative) shows the number of clips on Sep-28k-SW  labeled as 0 (absent) in Sep-28k but 1 (present) in Sep-28k-SW, and FP (false positive) are clips labeled as 1 (present) in Sep-28k but 0 (absent) in Sep-28k-SW. The percentage is calculated over total number of clips in Sep-28k-SW (2,621). Each clip can have multiple labels.}
\label{table::distribution}
\end{table}

In line with Sridhar and Wu's findings~\cite{charan2025whisper}, our analysis reveals substantial annotation disagreements between Sep-28k and Sep-28k-SW. Table \ref{table::distribution} shows the distribution of annotation disagreements across five types of stuttering events. While disagreements occur across all stuttering types, Sep-28k consistently labels more stuttering events than Sep-28k-SW -- resulting in more \textit{false positives} than \textit{false negatives} in Table~\ref{table::distribution}. This pattern highlights a significant gap in how stuttering is perceived and understood by PWS and non-PWS annotators. Our fine-grained dataset reviews and interviews with PWS AI professionals allowed us to identify four key factors that may contribute to this gap.

\subsubsection{Lack of Embodied Understandings of Stuttering}
A major source of the difference arises from a limited  understanding of stuttering by non-PWS. Both interview participants emphasized the importance of a contextual and embodied approach to annotation. Charan observed that a significant portion of the ``\textit{errors}'' in Sep-28k's labels stemmed from labeling ``\textit{clearly fluent speech}'' as stuttered:
\begin{quote}
``I think the labels [of Sep-28k] just kind of followed a rigid playbook. They didn't use any intuition. They had like a set of rules: If they hear `um,' it's an interjection. If a sound is dragged, it's a prolongation. If there's some pause, it's a block.''
\end{quote}
Rather than solely focusing on overt speech signals, he emphasized that stuttering should be perceived as situated within the flow of speech, noting that speakers may sometimes naturally prolong a word for non-verbal expression rather than due to stuttering:
\begin{quote}
    ``I feel like I can tell that `Oh, their voice is strained. That's stutter' versus `It's answering a question and there's a long dragged out `um.'' That's not stutter. It's just them thinking.''
\end{quote}
Echoing this quote, we observed many instances in Sep-28k where clips with natural disfluencies -- such as intentional interjections, prolonged interjection (e.g., \textit{ummmmm} to agree), pause because of thinking -- were labeled as stuttering.

Participants described how they drew on their embodied knowledge of stuttering when interpreting the data. Charan mentioned \textbf{accessory signals} such as \textit{``breathing, tone of speech, and speed of talking''}. Similarly, Rong explained the \textbf{kinematic signals} he used to identify  blocks, derived from his own sensation of stuttering: \textit{``You can hear the person is trying hard when they pause. Their teeth or vocal cords might produce small sounds.''} In another instance, when discussing a case that the lead author found difficult to distinguish between prolongation and block, Rong explained that prolongation ``\textit{should be continuous in the flow when pronouncing a sound},'' which is distinct from the choppy sounds associated with blocks.

Moreover, both participants noted the difficult case when PWS try to circumvent overt stuttering, yet they could better notice these ``tricks'' for masking stutters:
\begin{quote}
``If I practice properly, I can hide it [stuttering] pretty well. But I'll know where there was a small catch where no one else would notice.'' (Charan)
\end{quote}
\begin{quote}
``I can feel when they tried to replace the word. For example, they might pause at a place where non-stutterers typically wouldn't stop. They might speak fast but drag out the last word before he got stuck.'' (Rong)   
\end{quote}
While these less observable signals might feel intuitive to them, both emphasized that these perceptions are hard to translate into prescriptive rules. As Charan noted,
\begin{quote}
``The way I've focused on the breathing is very subjective. I couldn't describe a way to quantify breathing. Measures like speed of talking, the words per minute, are not a great way.''
\end{quote}
These subjectivities and the reliance on embodied knowledge make perceiving stuttering challenging for those without lived experience. The difficulty is further compounded when annotators must codify the subtle and fluid audio cues into written transcription.

\subsubsection{Challenges in Transcribing Disfluent Speech Verbatim}
The option to transcribe stuttered speech as it is is important to PWS~\cite{li2024want}, but significantly complicated the annotation process.
A notable challenge lies in distinguishing between sound-level and word-level repetitions, particularly with words that only have one syllable such as \textit{add}, \textit{how}, and \textit{what}, and when speakers speak fast. Rong highlighted the challenges of converting and preserving speech disfluencies into text transcriptions,
\begin{quote}
    ``You might need to listen to the clips many many times. You have to count the words they repeated. Sometimes they repeat very fast. In such cases, it's hard to get an accurate transcription and annotation even if you keep listening.''
\end{quote}

\subsubsection{Inherent Subjectivity in Speech Perception}
In some cases, participants recognized that speech perception is inherently subjective. Drawing from his experience training non-PWS annotators, Rong found that while annotations could be consistent in ``roughly 80\% of cases'', achieving complete agreement across all instances is ``impossible''. As he said, \textit{``The commonsense to everyone is different... It's impossible to make everyone the same, even when you annotate fluent speech.''} Echoing this quote, we observed many annotation differences in prolongations, rooted in annotators' subjective judgments of what prolongations sound like, particularly with vowel sounds in short words such as \textit{hear}, \textit{been}, \textit{ends}, \textit{oh}, \textit{two}, \textit{so}, and \textit{okay}. 

\subsubsection{Distortions Introduced by Audio Segmentation}
All of the above factors were further influenced by how the audio was segmented. As Sep-28k divides audio into strictly 3-second clips ~\cite{lea2021sep}, it is often difficult to determine stuttering events without access to the conversational flow and context. For example, segmentation can obfuscate whether a sound is part of a word or a disfluency caused by a block. In some cases, it is difficult to distinguish between words due to the cut, for instance, \textit{um} versus \textit{I’m}, or \textit{and} versus \textit{uh}. Charan thus suggested ``\textit{stringing together the clips instead of having 3-second clips.}''

\section{Co-Design Sessions}
Drawing on insights from our formative studies, we co-designed a set of annotation guidelines with PWS AI professionals and SLP professionals through a series of remote workshops conducted via Zoom between June and July 2025. To facilitate participants' thinking and discussions, we started with preliminary guidelines derived from our formative studies and worked with participants to test, critique, and iterate the guidelines through co-design sessions. Informed by our formative studies, we guided these sessions with three desirable qualities for annotations: (1) contextualized in speakers' speech flow; (2) prioritizing the PWS's embodied knowledge; (3) reflective of the trade-offs introduced by subjective speech perception. 

We began with PWS sessions to anchor in PWS's perspectives, followed by SLP sessions to integrate their professional perspectives. Overall, our co-design sessions sought to address two key questions when annotating stuttered speech:
\begin{itemize}
    \item \textbf{Ontology}: what labels should be included to represent stuttering events? 
    \item \textbf{Consistency}: where are annotation disagreements most likely to occur, and how to resolve them?
\end{itemize}

\subsection{Methods}
\subsubsection{Preliminary Guidelines}
We developed a set of preliminary guidelines to serve as a starting point for supporting and guiding discussions during our co-design sessions (available in the supplementary material). These guidelines were designed based on insights from our formative studies, co-author Wu's lived experience with stuttering, and a review of existing annotation guidelines and practices for stuttered speech, including FluencyBank~\cite{ratner2018fluency}, Sep-28k~\cite{lea2021sep}, and AS-70 ~\cite{gong202470}. The preliminary guidelines consisted of two components: (1) \textbf{an annotation framework for stuttering}, including the list and definition of stuttering events to be labeled, and (2) \textbf{general transcription practices}, such as how to transcribe accents and dialects, acronyms, and sensitive information. We decided it was crucial for the guidelines to support both verbatim and semantic transcriptions, as well as stuttering event detection, as these use cases are shown to be meaningful to the stuttering community~\cite{li2024towards, li-2024-codesign} and the development of disfluency friendly technologies~\cite{gong202470, charan2025whisper}.

We adopted the five stuttering events used in Sep-28k as our initial ontology, as these events represent the core disfluency types identified across prior work ~\cite{lea2021sep, li2024want, ratner2018fluency, gong202470}. We used the following markup for the five event types: \textbf{/r}: word or phase repetition; \textbf{/s}: sound repetition; \textbf{/b}: blocks; \textbf{/p}: prolongation; \textbf{/i}: interjection. The use of event markup allowed us to embed fine-grained stuttering event annotations within the text transcriptions. We also adopted elements from other annotation frameworks, such as FluencyBank ~\cite{ratner2018fluency}, to address the limits of Sep-28k (e.g. the lack of guidelines for speech transcriptions). The notations were designed to accommodate different downstream applications. For instance, we used brackets to indicate repeated segments, e.g., \textit{[pr-pr-pr-]/sprepare} (three repetitions of the \textit{pr} sound). Leaving the final ``\textit{pr}'' outside the brackets makes it easier to produce semantic transcriptions by removing all notations (e.g. ``\textit{/s}'') and bracketed content, while stripping the symbols (notations and [ ]) would generate verbatim transcriptions.  This scheme allows flexibility for different use cases. Finally, we reviewed established annotation practices for both stuttered and fluent speech ~\cite{lea2021sep, gong202470, li2024want, transcription-guide-online}, incorporating relevant components such as transcribing speech with accents and dialects, handling sensitive information, and representing numbers, symbols, and acronyms.

\subsubsection{Workshop Sessions with PWS AI Professionals}
We first conducted workshops with two PWS AI Professionals to center PWS's perspectives (Charan and Rong in Table \ref{table::participants}). We prioritized the perspectives of PWS in our discussions, while non-PWS members of our research team primarily served as observers and participated in annotation to support these discussions from non-PWS perspectives. We held three iterative sessions, each lasting between 40 and 60 minutes. Rong participated in one session, while Charan completed all activities.

Each session included the following activities: (1) First, PWS participants discussed the appropriateness and clarity of the annotation framework for stuttering in our preliminary guidelines, in comparison to other existing stuttering annotation frameworks.. These discussions helped to refine and familiarize the participants with our annotation framework for subsequent annotation activities. (2) Second, using the annotation framework, we annotated a 5-minute audio clip of stuttered speech we had collected. (3) Third, we compared our annotations and discussed the differences. We documented our insights (e.g., heuristics for identifying stuttering events) and iterated our guideline before the next session. We recorded videos of all sessions and compiled the artifacts developed, including document histories, for qualitative analysis. This analysis yielded themes on participants’ perceptions of the labeling framework, their suggestions, strategies for resolving differences, and views on annotation challenges.

\subsubsection{Workshop Sessions with SLP Professionals}
Following the workshops with PWS, we invited two SLP professionals (Jia/PWS, Kerrigan/non-PWS) with speech annotation experiences to further test and iterate on the guidelines. See Table~\ref{table::participants} for their backgrounds. The perspectives from SLP professionals are valuable, as they possess both clinical knowledge of stuttering and extensive experience in evaluating and annotating stuttered speech. As such, their participation could not only help assess the performance of our guideline but also enrich it with insights from a professional perspective. Jia and Kerrigan were also chosen for this study as they had  participated in stuttering advocacy and were well aware of the values and goals of the stuttering community. 

Given the time and attention required for annotation, we asked participants to complete their annotations prior to the session. We shared with them two audio clips of stuttered speech from our inclusive speech AI project (one 10-minute and one 5-minute clip), drawn from two participants who reported severe and moderate stuttering. We deliberately chose these samples to probe how the guidelines handle different levels of complexity in stuttered speech. To simulate the annotation process, we asked them to use Praat, a widely used speech annotation tool in SLP field~\cite{boersma2001speak} and for creating AI speech datasets~\cite{gong202470}. We imported machine (Otter.ai) generated transcripts into Praat as the starting point for annotation. Two researchers in our team reviewed the annotations by Jia and Kerrigan and highlighted the differences before the workshop. During the workshop session, we discussed these differences and gathered suggestions to refine our guidelines. The whole session lasted around 60 minutes. {We followed the same analysis approach as the sessions with PWS AI professionals. The analysis resulted in themes on issues participants pointed out, suggestions they provided, and their perspectives on the challenges in annotations.

\begin{table*}[!t]
\resizebox{\linewidth}{!}{
\begin{tabular}{p{0.09\linewidth}p{0.06\linewidth}p{0.19\linewidth}p{0.23\linewidth}p{0.43\linewidth}}
\hline
\textbf{Example} & \textbf{Session} & \textbf{Annotation 1}                     & \textbf{Annotation 2}                     & \textbf{Sources of Differences}                                                                                                       \\ \hline
\#1 & PWS & {[}A-A-{]}/sAdd                     & {[}A-a-{]}/s{[}add{]}/radd        
& number of repetitions; one-syllable words
\\ \hline
\#2 & PWS & {[}how{]}/r H/p/bow               & H/b How                            
& \begin{tabular}[c]{@{}l@{}} 
word repetition \textit{vs.} backtracking induced by block;\\
prolongation \textit{vs.} making sounds when encountering blocks
\end{tabular}
\\ \hline
\#3 & PWS & sh/bopping                     & [sh/p]/s shopping           & 
                                                
embodied conceptualization \textit{vs.} labeling based on audio cues
\\ \hline
\#4 & PWS & /b[O-O-]/sOpen              & [O-]/sOpen               &                                                           
a light block before ``\textit{open}''; number of repetitions
\\ \hline
\#5 & SLP & {[}Ha-ha-ha-ha-ha-ha-ha-{]}/shackathon & {[}ha-{]}/sha/p{[}ha-ha-ha-{]}/skathon &                                                               
number of repetitions; mixed stuttering events
\\ \hline
\#6 & SLP & fr{[}o-o-o-o-{]}/som                   & fro/pm                
& varied representations of sound repetition infused with elongation
\\ \hline
\#7 & SLP & w/porking                              & working                                        
& subjectivity in perception  of prolongation
\\ \hline
\end{tabular}}
\caption{Examples of differences in annotations surfaced in the sessions and their causes. Note that this list is not intended to be exhaustive but present typical patterns of differences.}
\label{table::difference}
\end{table*}

\subsection{Findings from PWS Sessions}

\subsubsection{Labeling Frameworks}
The PWS participants first discussed the labeling frameworks drawing on their own experiences with stuttering. For example, they questioned the necessity of FluencyBank's distinction between pauses within words and blocks before words, as both PWS participants perceived both as essentially the same experience. Two PWS participants also questioned FluencyBank's categorization of fillers (e.g., \textit{um}, \textit{you know}) as \textit{typical} disfluencies (i.e. not stuttering), noting that such interjections are often part of their stuttering experience. In general, the participants agreed with the stuttering event categorizations presented in the preliminary guidelines, and did not see the need to add or remove any labels from the framework.

Participants also discussed the value of labels and what information to encode in relation to potential use cases. In many cases, they weighed annotation granularity against the labeling workload and the technical feasibility of data processing. For example, while Rong noted the annotation of pause duration used in FluencyBank can be valuable to avoid speech models cutting PWS off, he added that this information can be inferred through algorithms if timestamps and block labels are available. Similarly, both PWS participants chose not to label typical disfluencies to reduce annotators' workload and prioritize tasks supporting stuttered speech. In the end, both participants agreed to adopt the five stuttering events used in Sep-28k as they offered a good balance between granularity and usefulness.

\BeginAccSupp{method=pdfstringdef,unicode,Alt={A screenshot of the Praat software interface which has five main parts. At the top of the interface, there is a text edit panel where users could put and edit texts for selected segment. Below the text edit panel, there is a large panel showing a horizontal audio waveform. A specific segment of the waveform is highlighted with a red box. Within this red box, a label "prolongation" points to the highlighted segment. Below the waveform, there are two panels. The top panel contains instructions, mentioning options to "zoom in to at most 10 seconds" and to "raise the 'longest analysis' setting." The bottom panel is a text editing area showing the transcription of the audio. At the very bottom of the screenshot, a progress bar indicates a range within the total duration.}}
\begin{figure*}[!t]
\centering
\includegraphics[width=14cm]{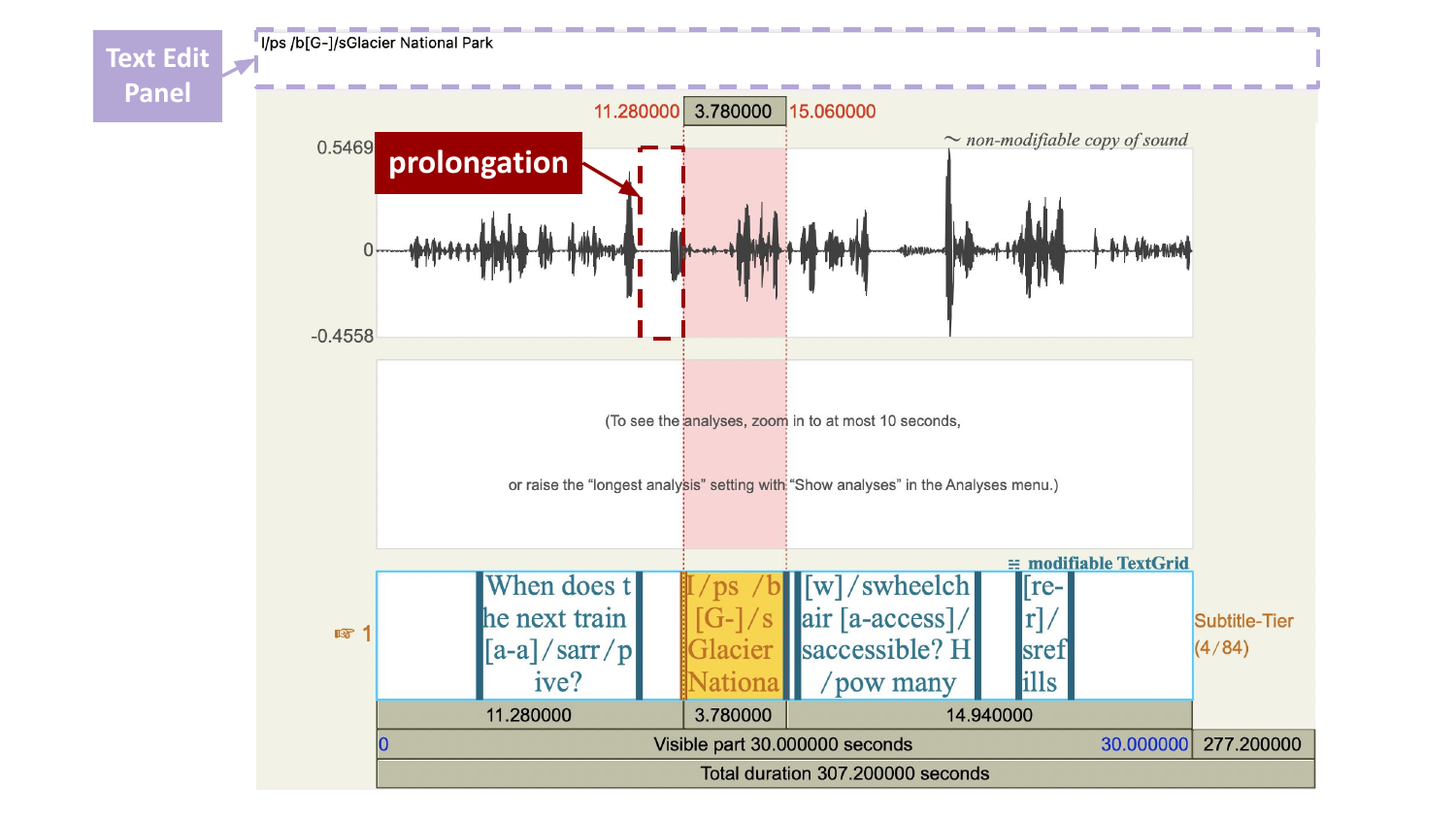}
\caption{{Screenshot of Praat interface, with a selected audio segment (in red) and corresponding editable text panel. The example shows how a prolongation might easily be overlooked due to selective listening and editing. The audio waveform corresponding to the prolongation labeled in \textit{I/ps} occurs outside the selected segment.}}
\label{fig::pratt}
\end{figure*}
\EndAccSupp{}

\subsubsection{Differences in Annotations and Resolving Strategies}
The annotation differences emerged during the sessions reflected challenges observed in our formative studies, e.g., difficulties in capturing numbers of fast repetitions, and identifying subtle or nuanced stuttering events. Table \ref{table::difference} presents examples of major differences we identified. Around these challenges, our discussion has led to three core principles for attending to potential differences in annotation: (1) centering PWS's embodied understanding; (2) grounding decisions in community interests; and (3) acknowledging subjectivity in speech perception.

\paragraph{Centering PWS's embodied understanding} A key way participants resolved differences was by sharing their embodied experiences of stuttering, as many disagreements stemmed from the gap between living with stuttering and observing it from the outside. Consider Example \#2 in Table \ref{table::difference} as an example. In this instance, the speaker appears to repeat the word ``how'' and elongate ``H'' in pronouncing the second ``how''. However, Charan did not label it as sound repetition or elongation, basing his judgment on his embodied experiences with stuttering. As he explained,
\begin{quote}
    ``I think it could be easily mistaken for a prolongation, because he was trying harder and he's still making sounds in the block. But I would still call that a block because he's not forced to drag it out. It's like he's trying a door multiple times. He's not pushing open the door slowly, like, a really heavy door that you'd want to push it open. A block, it's like a jam door, you have to, bump into it again and again.''
\end{quote}
In this instance, Charan based his interpretation on his embodied understanding instead of audio cues. He distinguished between intentional repetition and involuntary repetition caused by stuttering. In the context of stuttering, the former is known as backtracking ~\cite{heeman2006annotation}, a technique to pause and repeat a sound, word, or syllable to avoid an experienced block or disfluency. While the door metaphor Charan used vividly illustrated this distinction, this nuanced yet fundamental difference is hard to get without extensive exposure to stuttered speech or lived experience.

\paragraph{Grounding decisions in community interests} Participants also discussed the trade-offs involved in obtaining accurate annotations, given the significant human labor required. These discussions were grounded in their understanding of community interests. For example, when considering the challenges of counting repetitions, Rong emphasized the importance of taking into account the intended use of these annotations:
\begin{quote}
    ``It depends on what the models are trained for. You only need to get the event types right if your goal is to do stuttering event detection. However, if you wanna do transcription, you might need to get the number of repetitions right as well.''
\end{quote}
These discussions open up considerations of downstream applications and how annotation expectations may vary across contexts.

\paragraph{Acknowledging subjectivity in speech perception} In many cases, participants realized that speech perception is subjective in nature. A typical scenario is how to handle the omission of sounds in single-syllable words in speech flow. As Li (non-PWS co-author) observed,
\begin{quote}
    ``I feel we are consistent in what we heard. We all heard that there is a block before `\textit{is},' but it's hard to say it's the word `\textit{is}' repeated or the sound `\textit{i}' repeated.''
\end{quote}
In some cases, subjectivity was also evident in how stuttering was perceived across PWS, particularly in cases involving light or brief stuttering events. Consider the following discussion between Charan and Wu (PWS co-author) about a light block, for example:
\begin{quote}
    \texttt{Wu: It's hard for me to hear the block after the repetition.\\
    Charan: I just feel there's an unnecessary pause during the sound repetition. It’s not as rapid as sound repetition. There's like a pause there.\\
    -- They repeated the audio and listened to the segment again -- \\
    Wu: I just heard a lot of repetition of S. It's very hard to hear it [the block].}
\end{quote}
In this instance, the perception of a pause as `unnecessary' or `unnatural' is inherently subjective and difficult to articulate, often influenced by listeners' exposure to the speaker, embodied experiences, and judgments. These uncertain cases make an accurate representation of stuttering a nearly impossible goal to achieve. 

\paragraph{Summary} The sessions with PWS AI professionals provided rich insights that helped us refine our guidelines. The PWS rows in Table \ref{table::changes-guidelines} summarize changes across the sessions. A big takeaway from the discussions is the value of embodied knowledge that PWS bring to data interpretation. In response, we articulated and included key accessory and kinematic signals of stuttering participants shared, such as breathing patterns, speech flows, and vocal tension. However, the discussion also left many open questions, such as how to address the difficulties of converting speech flow into text and the inherent subjectivity with speech perception. To further explore the challenges in practice, we shared and tested these extended guidelines with SLP professionals in subsequent sessions.

\subsection{Findings from SLP Sessions}
The SLP professionals brought deep expertise grounded in their extensive exposure to stuttered speech and formal training. Their practice with our guidelines confirmed the challenges identified in earlier sessions, while also surfacing two additional ones that enhanced our understanding of the subjectivity in annotations: (1) the influence of audio processing within annotation software, and (2) the challenge of representing diverse stuttering patterns. They also offered suggestions to improve our guidelines and annotation process.

\subsubsection{Impact of Audio Processing Methods}
A key issue emerged concerned the effect of audio segmentation and processing in Praat, the annotation software widely adopted for clinical speech analysis as well as AI speech data annotation. In Praat, users select an audio segment based on timestamps and edit annotations in the text panel. While this design makes it convenient to listen to a segment repeatedly, the segmentation could influence how the speech was perceived. Even when we avoided breaking a word or a sentence, the timestamps provided by machine transcription services could still skip over subtle blocks or prolongations occurring just before or after a word or sentence as ASR models typically neglect non-speech content ~\cite{lamel-2002-trim} (see Figure \ref{fig::pratt} for an example). As Kerrigan explained,
\begin{quote}
    ``I'm just paying attention to the segments that have speech attached to them, and you need to be careful, because there will be segments that are not denoted as having speech in them. Then you go and you press that segment and you play it, and it's a block, or it's a prolongation. But it's just very, very quiet, and the mic is barely picking up on it.''
\end{quote}
Building on this point, Jia noted the variability in speech segmentation can affect how much context available to speech annotators and models to interpret stuttering. Even when annotators manually segment the entire conversation, the segmentation can still vary. Jia explained, as part of the SLP training, they have learned multiple standards for speech segmentation, and could apply them fluidly in different situations.   
Jia's point on segmentation variability sparked discussions on the potential impact of speech segmentation methods on models' ability to process stuttered speech. Since ASR models have increasingly relied on language models to produce most probable transcripts~\cite{radford2022robustspeechrecognitionlargescale}, how the speech is segmented would significantly influence the amount of context available to language models.

\subsubsection{Diverse and Intersecting Stuttering Patterns}
Another challenge lies in how to represent the multitudes of stuttering patterns. A common scenario involves labeling hybrid stuttering patterns. Although we categorized five stuttering event types, PWS often exhibit hybrid stuttering patterns, in which multiple types of stuttering occur simultaneously or overlap. Consider Example \#6 in Table \ref{table::difference} as an example. In this instance, while Jia labeled it as a prolongation, a more accurate way to describe what she heard was ``\textit{prolongation infused with repetition.}'' To her, how to represent these mixed and nuanced types of stuttering present a challenge when following a rigid guideline. 
\begin{quote}
    ``I do think sometimes even me being a person who stutters and hearing the stuttered speech, it's kind of hard to tell, if a block is infused with repetition, right? And sometimes it blocks with prolongation...Sometimes I don't want to over identify stuttering, but at the same time I don't want to not annotate them.''
\end{quote}
The quote above shows that even with lived experiences of stuttering, the judgment can still be rather subjective because of the heterogeneity of stuttering and the overlapping nature of stuttering events. As a result, both participants agreed that there is no single correct way to represent the speech but different perspectives. Both participants emphasized that differences in annotation should not be necessarily viewed as mistakes but as perspectives that can \textit{"enrich each coder’s perspectives",} as Kerrigan explained. Jia agreed, \textit{"we both did a good job, and we identified where the stuttered speech is, and it's important we're pretty consistent within ourselves. I think that's the best you can get from a coder."}

Although they annotated the speech differently, both believed that their approaches are grounded in deliberate considerations. Sometimes they might learn from the other version, while other times they might just respect their differences as part of the diversity of human speech perception.

\subsubsection{Suggestions for Improvement}
To better handle the complexity of annotation, participants offered valuable suggestions. First, noting that audio segmentation could cause annotators to miss stuttering moments, Kerrigan suggested reminding annotators to listen beyond the timestamped segments and to examine the actual audio waveform.

Second, participants cautioned against listing hard rules or heuristics, because these rules often add complexity to understand stuttering as an ``organic and complex'' phenomenon. Breathing, for instance, sparked discussions: while Charan had leveraged it to identify stuttering events,  Jia noted it can be both an avoidance technique and part of stuttering itself. She discussed the nuances with Wu (PWS co-author), as follows:
\begin{quote}
    \texttt{Jia: I would not pay too much attention to the breathing because that is almost like avoidance behavior, or people subconsciously do that to overcome those stuttering block. I feel this probably will add complexity to the already so complex work.}\\
    \texttt{Wu: But, as you said, if people are trying to avoid stuttering by using the breathing techniques, doesn't that mean that they are actually stuttering? Even if they could kind of cover it, it's actually a stutter.}\\
    \texttt{Jia: That's the one-million-dollar question, right? Because stuttering is so organic and complex. Sometimes trying not to stutter, is actually part of stuttering.}
\end{quote}
The conversation above reveals stuttering as a complex mix of involuntary disruptions and intentional bodily control, which blurs the boundary between stuttered and non-stuttered speech.

To preserve such multiplicities and allow room for interpretation, Wu proposed an evolving annotation model to replace the traditional majority voting method that assigns the most common label as the final label (e.g., ~\cite{lea2021sep}).
\begin{quote}
    ``We'll have one person code the whole thing, and then one person review their coding, and then one more person review their coding... We can always add more people that we think has more experience, or more knowledge in this, if we want to.''
\end{quote}
This idea gained support from both SLP participants, who further suggested strategies to better preserve embodied cues. For instance, Jia and Kerrigan both agreed that videos could be incorporated into the review process, as embodied cues such as lip movements are often a crucial aspect of stuttering and can aid interpretation ~\cite{tichenor2019stuttering}. 

\subsection{Evolution of Guidelines}
We integrated all participant feedback and updated the guidelines. The resulting guidelines are summarized in Section~\ref{sec::guidelines}, and their evolution is summarized in Table \ref{table::changes-guidelines}. In particular, we refined the definition of prolongation (from sound to syllable) to prevent confusion and account for non-speech sounds that might otherwise be overlooked in text annotations. We also added reminders about common mislabeling patterns, clarified ambiguous cases, incorporated participant-shared heuristics, and included notes on software usage. Crucially, we added notes to encourage reflexive practice and ensure that heuristics are not treated as prescriptive. 

Despite these efforts, the interpretation of stuttered speech proves to be inherently subjective and contextual, whether in navigating the messiness of speech flow, segmenting  audio, or interpreting stuttering events. To respect the speaker's ultimate authority in defining their speech patterns, we decided to gather feedback from PWS data contributors on annotations of their own speech and ground these evaluations in potential downstream applications. 

\section{Evaluation Sessions}
We recruited four PWS participants who had contributed their speech data and reviewed annotated samples of their recordings with them. Our sessions were guided by two questions:
\begin{itemize}
    \item How do PWS speakers perceive our labeling framework?
    \item How do PWS speakers perceive different versions of annotations?
\end{itemize}
The goal of this evaluation is not to provide a final assessment of performance but to explore PWS speakers' perceptions of our guidelines. Accordingly, we emphasized thematic saturation in recruitment and analysis.

\subsection{Methods}
Building on our earlier co-design sessions, we conducted one-hour Zoom interviews with four PWS speakers who contributed speech data to our inclusive speech AI initiative, Adedotun, Amina, Tatianna, and Benji. Their speech was annotated by at least two annotators familiar with stuttered speech, including annotations from annotators who did not participate in the co-design sessions. The annotators received the original audio recordings and TextGrid files containing auto-generated transcripts and timestamps produced by Zoom. Following the previous SLP session, the annotators used Praat to perform the annotation.

We received oral informed consent from the participants to conduct and record the interview for analysis. During the interviews, we asked participants about their perceptions of our annotation framework, the variations in annotations produced by different annotators, and their views on how such differences might affect downstream applications. We designed the following activities to guide these sessions. First, we presented examples of our annotated speech and asked the participants' general perception. We then presented different versions of annotations and asked participants about their perceptions and preferences. To facilitate reflection, we selected five examples of their annotated speech and invited participants to share their thoughts about the different versions.

To better contextualize our discussions, we asked our participants to consider the following five key scenarios and share their desired outcome and any potential issues with the annotations: (1) customizable auto-captioning in video conferencing, (2) voice commands for smart speakers, (3) speech-to-text features (e.g., for texting), (4) use in therapy contexts, (5) educational materials (e.g., training SLP students or promoting public awareness of stuttering). These scenarios were selected based on our PWS researcher's lived experiences with stuttering and our team's research expertise in this area. We also encouraged participants to suggest additional scenarios. To conclude the session, we brainstormed with participants about alternative ways to represent stuttered speech, inviting them to share any ideas they might have. We recorded all the video meeting sessions and transcribed the conversations for qualitative analysis. Our analysis identified themes consistent across participants regarding their perceptions of the labeling approach, annotation challenges, and label expectations.

\subsection{Findings}
Participants responded positively to our labeling framework. They noted that the labels capture a significant portion of stuttering (Adedotun, Tatianna), represent a big improvement over previous datasets (Benji), and emphasized the importance of verbatim transcription for reflecting a key aspect of their identity (Amina) and educating others about stuttering (Adedotun, Amina, Tatianna). However, their feedback also reflected limitations in relying on a fixed set of labels to capture their embodied experiences. This challenge fostered an appreciation for context-dependent evaluations of label accuracy.

\subsubsection{Limitations in Capturing Embodied Richness}
Participants brought complex embodied experiences that go far beyond what single labels can capture. In many instances, Adedotun and Amina identified more stuttering events in both annotations we presented, such as small interjections like ``\textit{uh}'', ``\textit{um}''. These short, light events could be easily missed by listeners but both participants were able to identity them due to
their familiarity with their own stuttering patterns. Participants also suggested additional heuristics based on their embodied experiences, citing cues such as changes in voice volume, speech rate, and the number of attempts made to push through moments of disfluency.

In many places, participants offered a nuanced interpretation that was hard for a single label to capture. For example, Amina was hesitant to apply the block label to an instance where she did not experience the block in her usual way:
\begin{quote}
    ``I don't know if it's a block per se, because I tend to interpret a block as something that happens before I pronounce a word, when I cannot pronounce a word at all. But here I was kind of able to pronounce the first part and then the second part later. So I don't know how to label this.''
\end{quote} 
Similarly, Benji highlighted the value of phonetic annotations, noting that people who stutter might ``\textit{transform sounds}'' during repetitions, making each repetition sound slightly different.

Participants also brought intimate understanding of their speech patterns that clarified the nuances missed in external observation. A common instance is that stuttering often blends tension and avoidance behaviors, making events easy to misclassify. For example, Amina explained that she often injects small interjections to get over blocks, which could blur the boundary between the two labels. Tatianna echoed similar points, attributing the differences in annotations to the gap between how stuttering sounds and how it is experienced internally:
\begin{quote}
    ``I don't fault annotation one, because it definitely sounded as if it was the repetitions. However, I know that's how I sound like when I have a block, as opposed to trying to bounce a sound.''
\end{quote} 
Furthermore, participants explained that they have certain stuttering patterns, which are important for differentiating labels. For example, Tatianna learned over years that her blocks are manifested mainly through repetitions and prolongations, while Benji described his stuttering as consisting primarily of repetitions. As stuttering manifests differently for each individual ~\cite{yairi2007subtyping}, participants interpreted and prioritized certain labels based on what aspects of their speech they find most meaningful. For example, Adedotun highlighted the value of labeling silences:
\begin{quote}
    ``The only label that I can think of [that could be added] is silence. Sometimes with stuttering, you're on a block, but there's no sound being produced.''
\end{quote}
Considering the dynamic nature of label meaning, our discussions then shifted to how speech labels should be interpreted in downstream applications.

\subsubsection{Situated Label Interpretation and Accuracy Expectations}
Participants expressed a desire to customize representations of their speech for different scenarios, including both verbatim and smoothed-over versions. Their expectations of accuracy shifted depending on conversational goals, social norms, relationships with interlocutors, and their own speech patterns. They shared many scenarios where they prioritized conveying content over representing stuttering events, e.g., when using voice commands with smart speakers, speech-to-text features, or talking to people who are familiar with their stuttering patterns such as family and friends. 

Participants also affirmed the value of verbatim transcription, particularly when speaking to those unfamiliar with stuttering. In such scenarios, including stuttering labels in transcriptions was seen as important for conveying personal identity and promoting stuttering awareness. Tatianna, for instance, recalled a past job interview where her stuttering was mistaken for a lack of competence. She wished her stuttering had been accurately documented so she could have evidence to advocate for herself. Adedotun explained how he imagined a stuttering-aware captioning system could help the listener better follow stuttered speech:
\begin{quote}
    ``[I] think for the person who's listening, it is helpful to know the time [of stuttering], so whenever the person is transitioning to the next word, they're aware that the stutter speech is either no longer happening, or is continuing to happen.''
\end{quote}

Participants expressed mixed expectations for therapeutic use, reflecting their differing orientations and goals. Some participants placed greater value on specific stuttering subtypes, as therapists need to ``\textit{learn from those different ways of stuttering and better take care of patients who stutter}'' (Adedotun), and to see their ``\textit{whole self}'' (Amina). Nevertheless, the utility of labels remains nuanced, as each individual may have their own perspective on what it means to speak `better' and it varies depending on contexts ~\cite{constantino2017rethinking}. For example, Adedotun and Benji placed less emphasis on specific stuttering events as they both prioritize self-acceptance in therapy. However, Adedotun acknowledged that for those who prioritize fluency, the specific labels might be important ``\textit{because the different ways of stuttering, you can use different techniques, so I think it would be important to have those nuances.}'' These comments highlight how the different utilities of labels and the varied orientations toward stuttering described by participants complicate the goal of annotation. Given the diversity within the stuttering community, we recognized that accounting for individual orientations and speech patterns presents an ongoing challenge for annotation design.

\subsubsection{Embracing Multiplicity with Care}
Participants demonstrated an understanding that human subjectivity is an inherent aspect of speech perception, provided the work is approached with care. Amina expressed a strong preference for having professionals handle the annotations:
\begin{quote}
    ``I would prefer they [the annotators] are at least familiar with the stuttered speech and know the different ways that stuttering occurs, not just a random person on the street.''
\end{quote}
This strong preference for professionals highlights annotation as a process of encoding human knowledge and perception, rather than just a mechanical or objective task. Echoing this point, Tatianna noted that speech perception is deeply shaped by listeners' personal language background, including potential differences between fluent and disfluent speakers, as well as between native and non-native English users.

Recognizing the inherent subjectivity in human perception, Adedotun valued the inconsistent labels, seeing them as a means to promote awareness and educate others about the diversity of human speech perception. Putting himself in the listener's position, he proposed ways to express ambiguity in speech perception by using mixed notations such as ``/b::/p, /b:/p, and /b=/p.'' These perspectives suggest that annotation is better understood as a form of collective sense-making, an ongoing effort to approach stuttering not as a fixed phenomenon, but as a multifaceted process requiring continuous reflection, interpretation, and care.

\subsection{Continuous Refinements After Evaluations}
While the PWS data contributors were generally satisfied with the annotations produced by our guidelines, the evaluation sessions highlighted a persistent gap between PWS's internal experiences and the observable signals perceived externally. This observation suggests that the annotation of stuttered speech requires an open-ended approach, allowing space for uncertainty and iterative refinements. We thus call for continuous involvement of community members and professionals as data stewards, with the power to quality control and update the annotations as our experiences and knowledge of stuttering evolve. At the time of writing, Kerrigan and Benji have assumed the stewardship role of reviewing and refining the annotations for all speakers. We anticipate, and are committed to, continuously revising our guidelines under the guidance of community data stewards.

\section{Annotation Guidelines and General Challenges}
\begin{table*}[!ht]
\begin{tabular}{p{0.25\linewidth}p{0.07\linewidth}p{0.62\linewidth}}
\hline
\textbf{Changes Made} & \centering{\textbf{Session}} & \textbf{Example} \\ 
\hline
Refined definitions & \centering{PWS} & Changed ``\textit{Elongated sound}'' to ``\textit{Elongated syllable}'' for the prolongation label definition. \\
\hline
Added reminders regarding common mislabeling 
&  \centering{PWS} & Added ``\textit{Be mindful of the potential difference between what you hear and the underlying stuttering events.}'' \\
\hline
Clarified ambiguous cases 
& \centering{PWS} &
Clarified ``\textit{For repeated short one syllable words that are hard to tell between sound vs word repetitions, we will just label them as word repetitions (unless it is very clear sound repetitions of the first sound).}''
\\ \hline
Included heuristics participants shared 
&  \centering{PWS} & 
Added ``\textit{Listen to non-verbal breathing sounds in front or end of the sentences.}''
\\ \hline
Added guidance on the annotation software use
&  \centering{SLP} & 
Added ``\textit{Listen beyond the timestamped segments and examine the audio waveform.}''
\\ \hline
Added clarifications for mixed stuttering events
& \centering{SLP} & 
Added a subsection titled ``\textit{multiple stutter}'' and provided examples to illustrate that ``\textit{one word can contain multiple stuttering events}''.
\\ \hline
Added reminders to encourage annotator reflections
& \centering{SLP} & 
Added ``\textit{Note that they are not prescriptive, but rather meant to guide deeper thought.}'' in the section of heuristics to identify stuttering events.
\\ \hline

\end{tabular}
\caption{Summary of guideline updates during co-design.}
\label{table::changes-guidelines}
\end{table*}

\label{sec::guidelines}
Here we summarize the annotation guidelines we developed. Due to  space limitation, the full guidelines are presented in Appendix \ref{appendix::guide}. We also show the major changes made through co-design sessions as we iteratively developed the guidelines. While our guidelines were derived from the lived experiences of stuttering, they are not intended to reduce or replace the involvement of PWS in the annotation process, but to recognize and highlight the expertise of disabled people in creating and annotating disability datasets. By sharing a list of inherent challenges in annotating stuttered speech, we argue for the necessity of active and continuous involvement of PWS in this process. 

\subsection{Annotation Guidelines Summary}

\subsubsection{Stuttering Annotation Guidelines} This section defines basic stuttering events to be annotated and introduces strategies to identify and distinguish them based on PWS expertise.

\begin{enumerate}
    \item \textbf{Basic Stuttering Events}
    \begin{itemize}
      \item Block (/b): a blocking pause before or within a word;
      \item Prolongation (/p): an elongated syllable;
      \item Sound repetition ([]/s): repeated sound (excluding single-syllable words);
      \item Word/phrase repetition ([]/r): repeated word or phrase, including single syllable word repetitions;
      \item Interjection ([]/i): common or individualized filler words.
    \end{itemize}

    \item \textbf{Mixed Stuttering Events}: the description of multiple stuttering events within one word (e.g. a sound is first repeated then prolongated) and how to annotate them, sometimes recursively. 

    \item \textbf{Identification Heuristics}: tips to capture the nuances in different stuttering events. For example, leveraging breathing sounds to detect small blocks; distinguishing prolongation and block based on airflow. Most heuristics were derived from the bodily experience of stuttering. Non-exhaustive and non-prescriptive, these heuristics are intended as a reminder to pay attention to non-verbal signals rather than hard rules.

    \item \textbf{Software Use}: examining the audio waveform rather than relying on the software's auto-segmentation, as sometimes stuttering events or cues can be cut out from auto segments.
\end{enumerate}

\subsubsection{General Speech Transcription Guidelines} This section describes transcription practices non-specific to stuttering.
\begin{itemize}
    \item \textbf{Consistency}: the transcription needs to match the speech verbatim, containing all the filler words, repetitions, and connecting words;
    \item \textbf{Special Cases}: various instructions on how to transcribe accents, dialects, numbers, symbols, and acronym, following the general rule of transcribing what is heard;
    \item \textbf{Private and Sensitive Information}: mark and redact sensitive information.
\end{itemize}

Table~\ref{table::changes-guidelines} shares major changes resulting from the discussions with different participants, providing insights into how the guidelines evolved throughout this study.

\subsection{General Challenges with Stuttered Speech Annotations}
Despite the comprehensiveness of our annotation guidelines, our findings also uncover inherent challenges in annotating stuttered speech:
\begin{itemize}
    \item The complex, organic, and often contested nature of stuttering. Producing faithful annotations thus requires embodied knowledge from PWS and, in some cases, insight from the speakers themselves.
    \item The ambiguity and subjectivity of human perception of speech flow. Experimental evidence shows that even subtle alterations to a stuffed toy presented to human subjects can shift their perception of vowel sounds~\cite{hay2010stuffed}.
    \item The socio-technical influences. Audio segmentation and the design of annotation tools can distort stuttered speech and omit key cues such as breathing or non-speech sounds.
\end{itemize}
We present a more detailed list in Appendix \ref{appendix::challenges} and share more examples in our supplementary files.

\section{Discussion}
Our work offers the first attempt to bring disability-first dataset practices into AI dataset annotation. While prior efforts of disability-first datasets often focus on the data collection process~\cite{sharma2023disability, theodorou2021disability, li2024want, gurari_vizwiz_2018, kamikubo2021sharing, massiceti2021orbit}, our study represents an important step in centering disabled people and their expertise throughout the AI development pipeline. Our guidelines differ from SLP frameworks developed to measure fluency and stuttering severity~\cite{ratner2018fluency, yaruss1997clinical}, as well as from earlier labeling frameworks designed primarily to detect stuttering events~\cite{lea2021sep}. Instead, our guidelines seek to capture the lived knowledge of stuttering from PWS. Next, we reflect on the lessons learned and discuss implications for stuttered speech datasets and accessibility datasets in general.

\subsection{Centering Disability Expertise in AI Dataset Annotation}
\label{sec::disability-expertise}
Our work demonstrates the importance of incorporating PWS's embodied knowledge in AI data annotation. Such knowledge represents what Hartblay called \emph{disability expertise} --``\textit{the particular knowledge that disabled people develop and enact about unorthodox configurations of agency, cultural norms, and relationships between selves, bodies, and the designed world}''~\cite{hartblay2020disability}. In our study, this expertise is reflected in both the embodied experience of stuttering and the awareness of community interests; both are essential for capturing stuttering and addressing annotation challenges, such as the need for situated evaluation. While stuttering can be particularly internalized and subjective, the importance of disability expertise in disability data annotation is shared with many other disabilities, as the  interpretation of data often relies on embodied knowledge unique to disabled bodies. For instance, hearing non-signers may overlook critical linguistic elements in sign language that deaf people are familiar with, such as facial expressions~\cite{erard_glove_reject, tang2026reimagining}. Sighted people often do not understand what information blind people want in visual descriptions~\cite{garg2025s, simons2020hope}. Neurotypical people may have a narrowed sense of what communication can look like for neurodivergent people, missing multisensory channels~\cite{alper2018inclusive, pinchevski2016autism}.

Missing disability expertise affects not only the accuracy of specific labels but also risks embedding ableism within the underlying AI training paradigm. For example, silent blocks are often mislabeled as non-speech segments and trimmed during ASR model training, leading to frequent interruptions of PWS speakers during speech blocks. Beyond stuttering, non-speech cues are central to disabled communicative practices~\cite{henner2023unsettling}, including those of autistic people ~\cite{alper2018inclusive}, deaf people ~\cite{kusters2017beyond}, and people with aphasia ~\cite{goodwin2004competent}. Ignoring the long-tail communication patterns and needs thus reflects a broader pattern of \emph{epistemic injustice} in AI ~\cite{ymous2020just, ajmani2024whose}. Fricker used the concept to describe the harm done to people in their capacity as a knower, excluding their perspective due to bias, prejudice, or structural inequality ~\cite{fricker2017evolving}. Historically, such epistemic injustice has led to decades of pushback from deaf communities against sign language technologies that misrepresent sign languages~\cite{erard_glove_reject, deaf-reddit}. Involving disabled people in AI data annotation should therefore be a central step to mitigate such injustice.

Moreover, the intimacy and complexity of embodied knowledge  pose deeper challenges for AI models to capture the full richness of disability. In the context of stuttering, Sheehan proposed the iceberg theory of stuttering ~\cite{sheehan1970stuttering}, noting that ``stuttering is like an iceberg, with only a small part above the waterline and a much bigger part below.'' From speakers' perspectives, stuttering involves more than observable behaviors; it also encompasses cognitive and affective reactions, including feelings, bodily sensations, and cognitive efforts that arise when anticipating moments of stuttering ~\cite{constantino2020speaker}. Reflecting similar ideas, the stuttering community advocates for more diverse evaluation measures, such as spontaneity rather than fluency ~\cite{constantino2020speaker}. The gap between what the current set of labels can capture and the richness of stuttering experiences reflects a long-standing question in HCI: the distinction between subjective, internal human experience and external observation ~\cite{leahu2008subjective}. Considering this gap, disabled people should have the authority to shape how their experiences are represented, interpreted, and evaluated in AI systems, a principle captured by the idea of disability-first. Next, we reflect on our practice of the disability-first principle and draw lessons for accessibility datasets more broadly.

\subsection{Toward Disability-First Annotations}
\label{sec::methods-reflection}
A key lesson we learned is the value of involving stakeholders with dense and diverse expertise. Prior research suggested recruiting disabled people is a key challenge in creating disability-first datasets~\cite{park2021designing}. The diversity of disabled experiences also makes exhausting population experiences nearly impossible~\cite{tang2025beyond, tang2024privacy}. In the context of stuttering, exposure to one's own stuttered speech may even trigger intense emotional responses~\cite{wu2023world}. Faced with these challenges, we carefully recruited participants who possess a wide range of expertise including embodied stuttering experiences, experiences as the organizers and advocates of the stuttering community, experiences with stuttered speech annotations, and technical proficiency. Still, we acknowledge that achieving a disability-first approach must remain an ongoing commitment, and we will continue involving community members and professionals in later annotation and stewardship work.

A further challenge lies in how to ensure sustained disability leadership in a mixed-ability research and development team, a common setting in related datasets efforts~\cite{theodorou2021disability, sharma2023disability, garg2025s}. Throughout our study, we prioritized the participation of PWS speakers in shaping our understanding of stuttering, with non-PWS members serving primarily in supporting roles. Our team ensured that PWS participants, including our team member, were given sufficient space and time to share their perspectives during team meetings or study sessions. When we encountered inconsistencies in annotations during the co-design sessions, PWS always voiced their perspectives first, while non-PWS focused on providing supporting evidence. Meanwhile, our non-PWS team members' positionality as non-stutterers, experienced UX researchers, and long-time disability advocates allowed them to ask targeted questions that helped the PWS participants articulate their feelings and sensation into scientific knowledge about stuttering. These clarifying questions from non-PWS members were instrumental to uncover deeper insights from the lived experiences and to practice reflexivity. All analysis results were reviewed by the whole team and verified by the PWS senior author to ensure we fairly represented PWS perspectives. We hope our experience can guide mixed-ability research teams to develop disability-first team dynamics in other tasks and settings.

While centered on stuttering, our disability-first approach has potential to inform the development of accessibility datasets more broadly. Similar to stuttered speech datasets, labeling issues are common in accessibility datasets and can benefit from the involvement of PWD~\cite{garg2025s, bragg2021fate}. For example, Garg et al. benchmarked vision-language models using a re-annotated subset of VizWiz, informed by a survey of blind participants~\cite{garg2025s}. The survey allowed them to incorporate information that blind people want but that sighted crowdworkers had not included~\cite{garg2025s}. Similarly, as explained in Section \ref{sec::disability-expertise}, many accessibility tasks require deeply personal embodied knowledge that disabled people possess, such as sign language translation~\cite{tang2023community, tang2026reimagining, bragg2021fate}, and disabled people's non-normative communication practices~\cite{alper2018inclusive, goodwin2004competent, henner2023unsettling}. Given the diversity of accessibility tasks, a key question for future work is to further understand the role of disability expertise in dataset annotation across different tasks. For instance, in tasks measured against WCAG guidelines, annotators may be considered qualified as long as they are familiar with those guidelines. However, disabled people should be involved to evaluate if WCAG is sufficient as the measure of accessibility~\cite{power2012guidelines}.

Future work should also investigate the challenges that arise in nuanced contexts when implementing the disability-first principle. In many tasks, disabled people might have limited access to sensory information that is important to the task~\cite{goodman2021toward, muehlbradt2022s}, or their desired information might come into tension with other values such as privacy or making racial inference from photos~\cite{bennett2021s, hanley2021computer, tang2025everyday}. In these situations, collaboration among people of mixed abilities and professionals is necessary. Crucially, while essential, the inclusion of disabled individuals is not a panacea~\cite{sloane2022participation, de2021good}, especially as the notion of `meaningful access' remains contested and evolving. For example, disabled people may lower their expectations due to the pervasive inaccessibility they have experienced in everyday life ~\cite{de2021good, tang2025everyday}. Disability-first is therefore more than just seeking user feedback; it involves equipping the communities with the resources, power, and support they need -- via training, advocacy, and sustained investment -- so that they can take ownership of the technologies that affect them.

\subsection{Embracing Multiplicity Throughout the AI Development Pipeline}
Ultimately, we advocate for acknowledging and working with the ambiguity and context dependencies of human data across the AI pipeline. Our findings show that some annotation challenges are inherent to human perception and cannot be fully resolved with a single guideline. These challenges echo recent work that treats data practices as reflexive practices ~\cite{cambo2022model} and frames inconsistencies as sites for reflection ~\cite{lin2023bias}. The case of stuttering provides yet another example showing that label inconsistencies are not always problems to be `settled'; rather, they reflect the diversity of human perception and can prompt ongoing, reflexive, and collaborative efforts to better understand human experiences. Similar to stuttering, many traditional accessibility tasks inherently involve ambiguity and subjective interpretation, such as sign language translation~\cite{bragg2021fate} and emotion recognition~\cite{begel2020lessons}. Subjectivity even appears in traditionally more `objective' tasks such as visual question answering (VQA)~\cite{tang2025everyday, ye2025semantic}. Although VQA has traditionally been treated as a task with a single ground-truth label~\cite{antol2015vqa}, recent work highlights the inherent subjectivity in visual information seeking, showing that blind users integrate multiple perspectives to shape their visual understanding, whether with AI tools or humans~\cite{tang2025everyday}.

Although task-specific research is needed, a multiplicity-aware perspective offers a valuable paradigm shift in data practices related to human perception. It calls researchers to be sensitive to practices erasing the diversity of human perception, such as rigid organizational control ~\cite{abdelkadir2025role} and majority voting~\cite{davani2022dealing}. Recent work has highlighted reflexive practices drawn from interpretivist traditions, such as encouraging data workers to reflect on their positionality~\cite{cambo2022model} and drawing on qualitative research methods to aid in data interpretation~\cite{wan2025noise}. Reflecting similar ideas, our SLP participants proposed a stewardship model in which a more experienced member oversees the annotation process. Building on this emphasis on interpretive labeling, tools could further support annotators in understanding differing perspectives, such as systems that predict the diversity of human responses~\cite{yang2018visual}. Such systems could also benefit users by revealing the different angles that informed labeling.

Specific to stuttering applications, a key lesson we learned is the situated meaning of speech labels. While PWS participants showed
strong intuition on which label `best' captures the speech patterns, 
they were also conscious about the space of interpretation and the fairness risk when deterministically assigning one label as the ``ground truth''~\cite{meyer2025perceptions}. The need to evaluate labels within the context of downstream applications has been highlighted by our participants.
For example, annotations might prioritize content accuracy in certain scenarios such as voice commands and speech-to-texting, while focusing  on accurate stuttering event labels in therapy or educational contexts. Additionally, PWS might prioritize the accuracy of certain labels according to their speech patterns. These varying priorities suggest that annotation should reflect the goals of the intended application rather than enforcing uniformity.

Another takeaway from our study is that interpretation of speech labels should extend beyond the moment of annotation. As PWS participants imagined the use of their annotated speech, the interpretation of labels is never final but continuously evolving; for example, inaccurate  annotations of stuttering type might still be useful in helping conversation partners become aware of the existence of stuttering, which could enhance listeners' comprehension of stuttered speech~\cite{Byrd-2017}. This way, the meaning of labels is co-constructed through their evolving use, rather than being fixed within datasets. Designing for such ongoing interpretations of labels requires approaches that encourage open, multiple interpretations ~\cite{sengers2006staying}. For example, systems could leave interpretations open to users using ambiguous labels such as /b::/p as Adedotun suggested. We can even envision a broader ecosystem that supports the ongoing collection of annotations from PWS. For instance, a self-annotation application could help PWS with self-therapy or communicate with SLPs, while the annotations could be shared to improve speech models or used for fine-tuning with informed consent. We encourage future work to explore continuous, participatory approaches that sustain annotation efforts over time and fully integrate diverse stuttering experiences into AI development pipelines.

\subsection{Limitations and Future Work}
This study has several limitations. First, our study relies on a relatively small number of participants. Despite our efforts in engaging multiple perspectives and community voices, our perspectives may be skewed toward people with higher socio-economic backgrounds and greater technical literacy, and under-represent those with multiple disabilities ~\cite{valente2025clinical}. Second, although the stuttering community has been exploring new ways to understand stuttering, our perspectives, including those of PWS, are inevitably influenced by pre-existing frameworks, such as those from SLPs. Part of our guidelines was still built off existing labeling frameworks. Given these two aspects, we argue that the guidelines should continue to evolve alongside people's experiences and perspectives. Future work should involve more community members to build upon and refine our efforts. Third, the speech data we collected and used did not cover variation across English dialects and other languages. Additional challenges could surface when extending our approach to more diverse speech patterns.

\section{Conclusion}
We present a case study exploring a disability-first approach to annotating AI datasets in the context of stuttered speech. Through a collaborative effort involving HCI researchers, PWS AI professionals, SLP professionals, and PWS data contributors, we explored the challenges of annotating stuttered speech and co-designed annotation guidelines grounded in the lived experiences of PWS. Our findings show the critical role of embodied knowledge in shaping more responsible annotation practices. We aim for this work to inspire broader efforts to center disabled knowledge and experiences throughout the AI development pipeline.

\begin{acks}
We thank all participants -- Charan Sridhar, Rong Gong, Jia Bin, Julia Kerrigan, Adedotun Bello, Amina Kobenova, Tatianna Vassilopoulos, and Benji Schussheim -- for their time and valuable insights. We are particularly grateful to our participants for consenting to the use of their real identities in this publication -- an act that reflects remarkable trust and confidence in our work.
We also thank the Authentic User Experience Lab at the University of California, Santa Cruz, especially Professor Norman Makoto Su, for stimulating discussions and collaboration on the future development of this work. Finally, we extend our heartfelt gratitude to all speech data contributors for entrusting us with their data that grounded this work. This work was supported by NSF Award \#2427710, the Patrick J. McGovern Foundation, and Borealis Philanthropy’s Disability Inclusion Fund. 
\end{acks}

\bibliographystyle{ACM-Reference-Format}
\bibliography{stuttering-annotation}
\appendix
\pagebreak
\section{Full Annotation Guidelines}
\label{appendix::guide}
In general, the speech need to be transcribed verbatim, meaning that, all repetitions, interjections (e.g. uh, wow, oh), filler words (e.g. you know, like), and even stuttered sounds need to be transcribed in the text as they are. The following sections will break down the specific requirements and standards for transcription.\\\\
{\large\textbf{1. Stuttering Annotation}}\\
{\large\textbf{1.1 Basic annotation}}\\ 
For stuttering or other non-stuttering disfluencies such as self-correction or repeated words, transcribe what you heard verbatim. When the disfluency is induced by stuttering, also mark the type of stutter using the following annotations described in Table~\ref{tab::annotation_types}.

\begin{table*}[h]
\resizebox{\linewidth}{!}{
\begin{tabular}{|l|l|l|l|}
\hline
\textbf{\begin{tabular}[c]{@{}l@{}}Stutter type\end{tabular}} & \textbf{Code} & \textbf{Example}                                                                                                                                         & \textbf{Notes}                                                                                                                                                                                                                                                                                                                                                                                                                                                                                                                                       \\ \hline
Block                                                            & /b            & \begin{tabular}[c]{@{}l@{}}- My /bname (blocking on “n”)\\ - Spa/bghetti  (blocking on “g”)\end{tabular}                                                 & \begin{tabular}[c]{@{}l@{}}A blocking pause before or within a word. \\ Insert /b right before the next sound/word. \\ While not being prescriptive, to attend to \\ these blocks, you can pay attention to the \\ airflow, a gush of air right before a sound \\ (e.g. a strong airy “t” sound in “turn”) \\ can indicate a block, even when there \\ is no clear pause in speech.\end{tabular}                                                                                                                                                     \\ \hline
Prolongation                                                     & /p            & \begin{tabular}[c]{@{}l@{}}M/pommy\\ Mommy, with “m” sound elongated.\end{tabular}                                                                       & \begin{tabular}[c]{@{}l@{}}Elongated syllable. Insert /p right after the\\ prolongated syllable\end{tabular}                                                                                                                                                                                                                                                                                                                                                                                                                                         \\ \hline
Sound repetition                                                 & {[} {]}/s     & \begin{tabular}[c]{@{}l@{}}{[}pr-pr-pr-{]}/sprepare\\ Prepare, with “pr” repeated three times. \\ Verbatim transcript:\\ “pr-pr-pr-prepare”\end{tabular} & \begin{tabular}[c]{@{}l@{}}Repeated syllables. Transcribe the exact number \\ of times the sound/syllable is repeated, put a “-” \\ after each one and put all extra text in brackets, \\ insert the /s right after the brackets.\end{tabular}                                                                                                                                                                                                                                                                                                       \\ \hline
Word/Phrase repetition                                           & {[} {]}/r     & {[}my my{]}/r my name                                                                                                                                    & \begin{tabular}[c]{@{}l@{}}The same word or phrase is repeated. Transcribe \\ the exact words/phrases as repeated, put the extra \\ parts in brackets with space in between each word, \\ insert /r right after the bracket, leave a space between \\ /r and the next word.\end{tabular}                                                                                                                                                                                                                                                             \\ \hline
Interjection                                                     & {[} {]}/i     & I {[}uh{]}/i work                                                                                                                                        & \begin{tabular}[c]{@{}l@{}}Common filler words such as "um" or "uh" or \\ person-specific filler words that individuals use \\ to cope with their stutter (e.g., some users frequently \\ say "you know" as a filler). Transcribe the \\ stuttering-related filler words as they are \\ and put them in the brackets, followed by the /i mark. \\ Leave a space between /i and the next word.\\ * You need to transcribe non-stuttering filler \\ words verbatimas well, but do NOT use \\ {[} {]}/i to mark them as stuttering events.\end{tabular} \\ \hline
\end{tabular}
}
\caption{Basic stutter types and corresponding annotation markups}
\label{tab::annotation_types}
\end{table*}

\noindent\textbf{Notes:}

\textbf{1.2 Multiple stutter}\\
One word can contain multiple stuttering events, thus be annotated with more than one stuttering code. A few examples:
\begin{itemize}
    \item ``[m-m-m-]/sm/py'':  the ``m'' sound in ``my'' is repeated three times and prolongated;
    \item  ``I [uh uh uh]/r/i /bwork'': ``uh'' as a stuttering interjection was repeated three times, and followed by a block before the speaker said ``work''.
\end{itemize}

\textbf{1.3 Single syllable word repetition}\\
For repeated short one syllable words that are hard to tell between sound vs word repetitions, we will just label them as word repetitions (unless it is very clear sound repetitions of the first sound).

\textbf{1.4 Be mindful of the potential difference between what you hear and the underlying stuttering events.}\\
Some heuristics our team observed in our pilot stage are listed below. Note that they are not prescriptive, but rather meant to guide deeper thought.

\begin{enumerate}
    \item  Listen to non-verbal breathing sounds in front or end of the sentences. This is particularly important for detecting small blocks.
    \item Distinguish involuntary repetition and block then backtracking. Sound repetition is involuntary but the repetition from block and backtracking is intentional. We should label it as a block.
    \item Distinguish prolongation vs block and trying to make the sound during block. The airflow is continuous in prolongation, but choppy in blocks. If the air is not flowing, we should label it as a block.
\end{enumerate}

\textbf{1.5. listen beyond the timestamped segments and examine the audio waveform. Sometimes stuttering events or cues could be missed outside of the segments.}\\

{\large\textbf{2. Consistency}}\\
The transcript needs to be consistent with the speech, even if the speaker makes grammatical errors. Keep all the connecting words such as ``and'', ``but'', and ``so''. 

{\large\textbf{3. Accents and Dialects}}\\
Do not transcribe accents or dialects by pronunciation unless there is a common spelling for the accented word (e.g. ``y'all''). Use contractions only if they are used by the speaker. If the speaker mis-pronounced a word, use the intended word in the transcription, except when the speaker mis-spoke the entire word. 

For example, when the speaker intended to say ``royal palace'' but said it as ``royal place'', you should transcribe it verbatim as ``royal place''. But if the speaker said ``royal plaze'', you should transcribe it as ``royal place''.

{\large\textbf{4. Sensitive Information}}\\
Mark sensitive information, such as names, age, specific occupation, specific places, in < >. For example: I am <Katie> and I live at <Athens, Ohio>.

{\large\textbf{5. Numbers and Symbols}}

Transcribe numbers in English rather than Arabic numbers. E.g. ``We will meet at two thirty''.

Symbols should be transcribed by the sound, too. E.g., when the speakers mention an email address or website, it should be transcribed as ``info at university dot org'', not ``info@university.org''. 

{\large\textbf{6. Acronyms}}

Transcribe acronyms as it was said, for example, NASA will be transcribed as ``NASA'', not ``N A S A''. 

If the acronym is spelled out letter by letter to emphasize the spelling, add a hyphen in between letters. For example, ``I am part of the A-C-T, ACT program.'' 

If it is a common acronym that is not spelled out by the speaker, write it without hyphen. E.g. ``I have ADHD”, or ``AI is changing the world''.

\section{Challenges in Annotations}
\label{appendix::challenges}
This list is not prescriptive or intended to be exhaustive. The main goal to list these notes and thoughts is to spark future reflections and discussions over the challenges in annotating stuttered speech. Please refer to our supplementary files for a more detailed list and related example clips.

\begin{enumerate}
    \item Lack of Embodied Understandings of Stuttering
    \begin{itemize}
        \item Annotators didn't differentiate stuttering events and `natural' or intentional disfluencies.
        \item Annotators didn't understand the labels as how PWS feel about them.
        \item Speakers' internal experience of their speech often differs from what listeners are able to perceive.
    \end{itemize}
        \item Diverse and Intersecting Stuttering Patterns
    \begin{itemize}
        \item Hard to objectify stuttering events because of the diverse speech patterns, e.g., many instances resemble mixed events, PWS may draw on techniques to manage stuttering.
    \end{itemize}
    \item Challenges in Transcribing Disfluent Speech Verbatim
    \begin{itemize}
        \item Repeat fast and repeat many times, e.g., \textit{[n-n-n/p-n-]/snavigate to mom's house}.
        \item Sound repetition vs. word repetition for single-syllable words, e.g., \textit{[Aa]/s [add]/r add a pocket projector for Katie to my gift list.}
    \end{itemize}
    \item Inherent Subjectivity in Speech Perception
    \begin{itemize}
        \item The perception of elongation of vowels is inherently subjective especially without listening to the speech flow, and for long vowel and compound vowel: \textit{hear, been, language, oh, ends, two, so, okay, wow}.
    \end{itemize}
    \item Distortions Introduced by Audio Segmentation
    \begin{itemize}
        \item Hard to contextualize the sounds because of the audio segmentation.
        \item Hard to tell the words because of audio segmentation, e.g., \textit{um} vs. \textit{I’m}, \textit{and} vs. \textit{uh}.
    \end{itemize}
    \item Impact of Audio Processing Methods
    \begin{itemize}
        \item Easy to miss non-speech content because of how Pratt renders segments; annotators might click the segment while missing cues outside of the segments.
    \end{itemize}
\end{enumerate}

\section{Interview Guide Used in Formative Studies}
\label{appendix::interview-guide}
\begin{enumerate}
    \item How would you describe stuttering to people who do not stutter?
    \item What would you want AI models to learn about stuttering?
    \item Can you walk me through a typical workflow you followed when annotating the speech data?
    \item What differences did you observe between the labels you assigned and those from crowdworkers?
    \begin{itemize}
    \item  How did you feel about these differences?
    \item  What factors do you think might cause these differences?
    \end{itemize}
    \item I picked some examples from the dataset you annotated. Can you walk me through how you did annotations for these examples?
    \begin{itemize}
        \item Did you feel anything hard to make a decision during the process? 
        \item Do you remember any other hard cases? What makes these examples hard?
    \end{itemize}
    \item If you have an opportunity to re-design the annotation process, how would you design the annotation task?
    \begin{itemize}
        \item What labels/features would you design to represent stuttering?
        \item What expertise do you think PWS or non-PWS have in annotations? 
    \end{itemize}
\end{enumerate}
\end{document}